\documentclass[a4paper,11pt]{article}
\usepackage{jheppub} 
\usepackage{lineno}

\arxivnumber{2402.06736} 
\usepackage{amsmath}
\usepackage{amsfonts}
\usepackage{hyperref, url}
\usepackage{graphicx,subfigure}
\usepackage[normalem]{ulem}
\usepackage{epsfig}
\usepackage{color,soul}
\usepackage{multirow}
\usepackage{array}
\usepackage[utf8]{inputenc}
\usepackage[T1]{fontenc}
\usepackage[dvipsnames]{xcolor}
\usepackage{comment}
\usepackage{aas_macros}
\hypersetup{colorlinks   = true,
            urlcolor     = blue,
            citecolor    = blue,
            linkcolor    = blue,
            menucolor    = blue,
            anchorcolor  = blue,
            filecolor    = blue}
\enlargethispage{\baselineskip}

\usepackage{graphicx}

\newcommand{\Mpl}{M_{\text{pl}}}

  \def\l{\lambda}

 \newcommand{\Ocal}{{\mathcal O}}

\begin{document}
\preprint{KEK-QUP-2024-0002, KEK-TH-2599, KEK-Cosmo-0338, IPMU24-0004}

\title{Bosenovae with Quadratically-Coupled Scalars in Quantum Sensing Experiments}


\author[a,b]{Jason Arakawa}
\author[a]{Muhammad H. Zaheer}
\author[c,d]{Joshua Eby}
\author[b,d,e,f]{Volodymyr Takhistov}
\author[a,g]{Marianna S. Safronova}

\affiliation[a]{Department of Physics and Astronomy, University of Delaware, Newark, Delaware 19716, USA}
\affiliation[b]{International Center for Quantum-field Measurement Systems for Studies of the Universe and Particles (QUP), KEK, 1-1 Oho, Tsukuba, Ibaraki 305-0801, Japan}
\affiliation[c]{The Oskar Klein Centre, Department of Physics, Stockholm University, 10691 Stockholm, Sweden}
\affiliation[d]{Kavli Institute for the Physics and Mathematics of the Universe (WPI), The University of Tokyo Institutes for Advanced Study, The University of Tokyo, Kashiwa, Chiba 277-8583, Japan}
\affiliation[e]{Theory Center, Institute of Particle and Nuclear Studies, High Energy Accelerator Research Organization (KEK), Tsukuba 305-0801, Japan}
\affiliation[f]{Graduate University for Advanced Studies (SOKENDAI),
1-1 Oho, Tsukuba, Ibaraki 305-0801, Japan
}
\affiliation[g]{Joint Quantum Institute, National Institute of Standards and Technology and the University of Maryland, College Park, Maryland 20742, USA}

\emailAdd{arakawaj@udel.edu, hani@udel.edu, joshua.eby@fysik.su.se, vtakhist@post.kek.jp, msafrono@udel.edu}

\date{\today}

\abstract{
Ultralight dark matter (ULDM) particles of mass
$m_\phi \lesssim 1~{\rm eV}$ can form boson stars in DM halos. Collapse of boson stars leads to explosive bosenova emission of copious relativistic ULDM particles. In this work, we analyze the sensitivity of terrestrial and space-based experiments to detect such relativistic scalar ULDM particles interacting through quadratic couplings with Standard Model constituents, including electrons, photons, and gluons. We highlight key differences with searches for linear ULDM couplings. Screening of ULDM with quadratic couplings near the surface of the Earth can significantly impact observations in terrestrial experiments, motivating future space-based experiments.
We demonstrate excellent ULDM discovery prospects, especially for quantum sensors, which can probe quadratic couplings orders below existing constraints by detecting bosenova events in the ULDM mass range $10^{-23}\,{\rm eV} \lesssim m_\phi \lesssim 10^{-5}\,{\rm eV}$. We also report updated constraints on quadratic couplings of ULDM in case it comprises cold DM.}

\maketitle
\flushbottom
\section{Introduction}

The predominant component of matter in the Universe, contributing $\sim$85\%, resides in mysterious dark matter (DM)~(see e.g. Ref.~\cite{Cooley:2022ufh} for an overview).
Despite decades of effort, DM continues to evade direct detection aiming to detect DM energy depositions in experiments due to non-gravitational DM interactions.
Significant attention has been devoted to exploring weakly interacting massive particles (WIMPs),  
which have typical masses in the range of $\sim 1$ GeV – 100 TeV (see e.g.~Ref.~\cite{Roszkowski:2017nbc,Schumann:2019eaa} for reviews). A well-motivated class of DM that has been less studied is ultralight dark matter (ULDM) \cite{ULDM}, having mass
$m_{\phi} \lesssim 1\,{\rm eV}$. With such small $m_{\phi}$, the DM is characterised by very large occupation numbers and behaves like a coherent classical wave~(e.g.~\cite{Hu:2000ke,Hui:2016ltb}). The associated direct DM detection signatures are also distinct.

In the case of scalar ULDM, direct couplings with the Standard Model (SM) can lead to the variation of fundamental constants, such as the fine-structure constant $\alpha$ \cite{Arvanitaki:2014faa, PhysRevLett.115.201301}. 
The experimental program searching for scalar ULDM has particularly benefited from variety of approaches based on high-precision instruments~\cite{ULDM,Antypas:2022asj,adams2023axion}, such as atomic clocks \cite{PhysRevLett.117.271601, Beloy:2020tgz, Filzinger:2023zrs, Sherrill:2023zah, 2022PhRvL.129x1301K, Hees:2016gop, Arvanitaki:2014faa, Aharony:2019iad, Banerjee:2023bjc, SrOH2021, Zhang:2022ewz, Zaheer:2023ulf}, optical cavities \cite{Kennedy:2020bac, Tretiak:2022ndx, Campbell:2020fvq, Geraci:2018fax}, spectroscopy \cite{Oswald:2021vtc, VanTilburg:2015oza}, mechanical resonators \cite{Branca:2016rez, Arvanitaki:2015iga, Manley:2019vxy}, optical interferometers \cite{PhysRevLett.114.161301, PhysRevA.93.063630, Grote:2019uvn, Aiello:2021wlp, Vermeulen:2021epa, Fukusumi:2023kqd}, atom interferometers \cite{Badurina:2021rgt, 2021QS&T....6d4003A}, tests of the equivalence principle (EP) \cite{Hees:2018fpg,Berge:2017ovy}, and fifth force searches \cite{Fischbach:1996eq, Adelberger:2003zx, Murata:2014nra} whose complementary sensitivities span and cover a wide ULDM mass range. 

ULDM can form gravitationally-bound boson stars~\cite{Kaup:1968zz,Ruffini:1969qy,Colpi:1986ye} through both gravitational interactions~\cite{Levkov:2018kau,Chen:2020cef} and 
the self-interactions of the ULDM field~\cite{Kirkpatrick:2020fwd,Chen:2021oot,Kirkpatrick:2021wwz}. In the presence of attractive self-interactions, the boson star is expected to become unstable once it reaches a critical mass. This occurs when the density of the boson star increases enough for the ULDM self-interactions to destablize the previously-established hydrostatic equilibrium between the boson star's gravitational and the gradient energy~\cite{Chavanis:2011zi,Chavanis:2011zm,Eby:2014fya}. Subsequently, the boson star collapses and, due to the same self-interactions, explodes in a \textit{bosenova} event, copiously emitting relativistic ULDM scalars~\cite{Eby:2016cnq,Levkov:2016rkk,Helfer:2016ljl,Eby:2017xrr}. Due to wave-spreading of relativistic scalar bursts in-flight away from the source the signal can persist for a significant time, even months to years, as it traverses terrestrial or space-based experiments~\cite{Eby:2021ece}. In previous work \cite{Arakawa:2023gyq}, we explored transient signals from bosenovae, establishing prospects for enhanced reach in case of dilatonic ULDM couplings to SM present at first order in the effective field theory (EFT). 
Some of us also have explored the prospects for the detection of bosenovae from axion stars~\cite{Eby:2021ece}. Complementarily, relativistic particles originating from historic bosenova events can contribute to diffuse background with distinct observable signatures~\cite{Eby:2024mhd}.

In this work we explore the detection prospects for bosenovae focusing on quadratic couplings of ULDM to the SM.  
Quadratic couplings could dominate over linear ULDM couplings in scenarios with additional symmetries, such as a $\mathbb{Z}_2$ symmetry under which ULDM $\phi$ field is odd.
This is readily realized for axions, which transform as a pseudoscalar field under parity symmetry. As a consequence, axion-like fields including the QCD axion generically possess quadratic couplings without corresponding linear ``dilatonic'' coupling~\cite{Kim:2022ype,Beadle:2023flm,Kim:2023pvt}. In contrast to the analysis of observational bosenova signatures with linear ULDM couplings to SM~\cite{Arakawa:2023gyq}, quadratic interactions posses additional phenomenological consequences, such as screening of the field in the presence of ordinary matter~\cite{Hees:2018fpg}.

The paper is organized as follows. First, in Sec.~\ref{sec:EFT} we describe the EFT characterizing the ULDM-SM interactions, as well as the self-interaction coupling. Then, in Sec.~\ref{sec:bosonstar} we give an overview of boson stars and bosenovae which provide the astrophysical signal. Next, in Sec.~\ref{sec:detection} we give an outline of the detection methods for bosenovae and experimental sensitivities. In Sec.~\ref{sec:results} we describe implications of our results. We conclude in Sec.~\ref{sec:conclusions}.

\section{Theoretical Model}
\label{sec:EFT}
\subsection{Effective interactions}

A real scalar field $\phi$ can be generally characterized by a Lagrangian
\begin{align} \label{eq:Lphi}
    \mathcal{L}_\phi &= \frac{1}{2}\partial_{\mu}\phi \partial^{\mu} \phi - \frac{1}{2}m_{\phi}^2 \phi^2 
    + \frac{\lambda}{4!}\phi^4 + ...
\end{align}
where $m_\phi$ is the field's mass and $\lambda$ is the self-interaction coupling, and the $[\dots]$ indicate possible terms with higher powers of $\phi$. Here, we consider $\phi$ to constitute ULDM with mass $m_{\phi} \lesssim {\rm eV}$ and treat $|\lambda|\ll 1$ as a free parameter that depends on the underlying theory. We focus on parameter space $\lambda>0$, denoting attractive self-interactions between $\phi$ particles\footnote{This is also motivated by some theoretical considerations, see e.g.~\cite{Fan:2016rda}.}.
In Sec.~\ref{ssec:lambdaconstraints} we discuss various constraints on $\lambda$.

The couplings of $\phi$ with the SM, and as a consequence their optimal detection methods, will depend on the symmetries such as the parity of the model. Axions and axion-like particles more generally (see e.g.~\cite{adams2023axion}) are typically parity-odd, which implies derivative couplings to SM fields. On the other hand, parity-even fields (see e.g.~\cite{Antypas:2022asj}) have different interactions with SM fields. As we discuss below, this leads to distinct detection prospects and promising search strategies.

We can generally characterize the possible dilatonic-type (i.e. parity-even) couplings of $\phi$ to the SM
by their EFT Lagrangian contributions
\begin{align} \label{eq:Lint_EFT}
    \mathcal{L}_{\text{int}} = \sum_{n,i} d^{(n)}_i \bigg(\frac{\sqrt{4\pi}\,\phi}{\Mpl}\bigg)^n \mathcal{O}^i_{\text{SM}}~,
\end{align}
where $\Mpl=1.2\times 10^{19}\,{\rm GeV}$ is the Planck mass and $\Ocal_{\rm SM}^i$ are dimension-four SM operators labeled by index $i$. In this parametrization the $d_i^{(n)}$ couplings are dimensionless. 
Throughout, we will focus on scalars with quadratic $\propto \phi^2$ interactions with SM constituents, corresponding to Eq.~\eqref{eq:Lint_EFT} with $n=2$. In particular, we consider
\begin{align} \label{eq:Lint}
    \mathcal{L}_{\text{int}} \supset &~\frac{4\pi\,\phi^2}{\Mpl^2}
    \bigg(  d_{m_e}^{(2)}m_e \overline{e} e 
        - \frac{d_{e}^{(2)}}{4} F_{\mu \nu}F^{\mu \nu}
        - \frac{d_g^{(2)} \beta(g_s)}{2g_s}G_{\mu\nu}^a G^{a\mu\nu}\bigg)~,
\end{align}
where $d_{m_e}^{(2)}$, $d_{e}^{(2)}$ and $d_g^{(2)}$ are couplings of $\phi^2$ to SM, $F_{\mu\nu}$ is the electromagnetic field strength tensor, $G_{\mu\nu}^a$ is the quantum chromodynamics (QCD) field strength tensor, and $e$ is the electron field with mass $m_e$. The gluon interactions depend on the ratio $\beta(g_s)/g_s$ of the QCD beta function to the strong coupling.

In the presence of the couplings described in Eq.~\eqref{eq:Lint}, classical oscillations of the ULDM field at frequency of order $m_\phi$ lead to oscillations of the fundamental constants, $m_e/m_p$ (electron-to-proton mass ratio), $\alpha$ (fine structure constant), and $m_q/\Lambda_{\rm QCD}$ (ratio of quark mass to QCD scale), 
respectively. As we discuss, this allows for unique opportunities to search for their manifestations in quantum sensing experiments. 
Typically, such searches focus on detecting the local DM abundance in the vicinity of Earth, which is expected to have an energy density of order $\rho_{\rm DM}\simeq 0.4\,{\rm GeV/cm}^3$~\cite{Read:2014qva}. However, burst emissions of relativistic ULDM arising from Galactic transient sources establish novel and unique complementary observational targets~\cite{Eby:2021ece,Arakawa:2023gyq}. 
Previously~\cite{Arakawa:2023gyq}, we analyzed the detection prospects for ULDM with linear $\propto \phi $ interactions with the SM constituents, corresponding to $n=1$ case in Eq.~\eqref{eq:Lint_EFT}. In this study, we analyze the signatures and detection prospects for ULDM scalars with quadratic SM couplings.

\subsection{Screening}
\label{sec:Screening}

\begin{figure*}
    \centering
    \includegraphics[width = \linewidth]{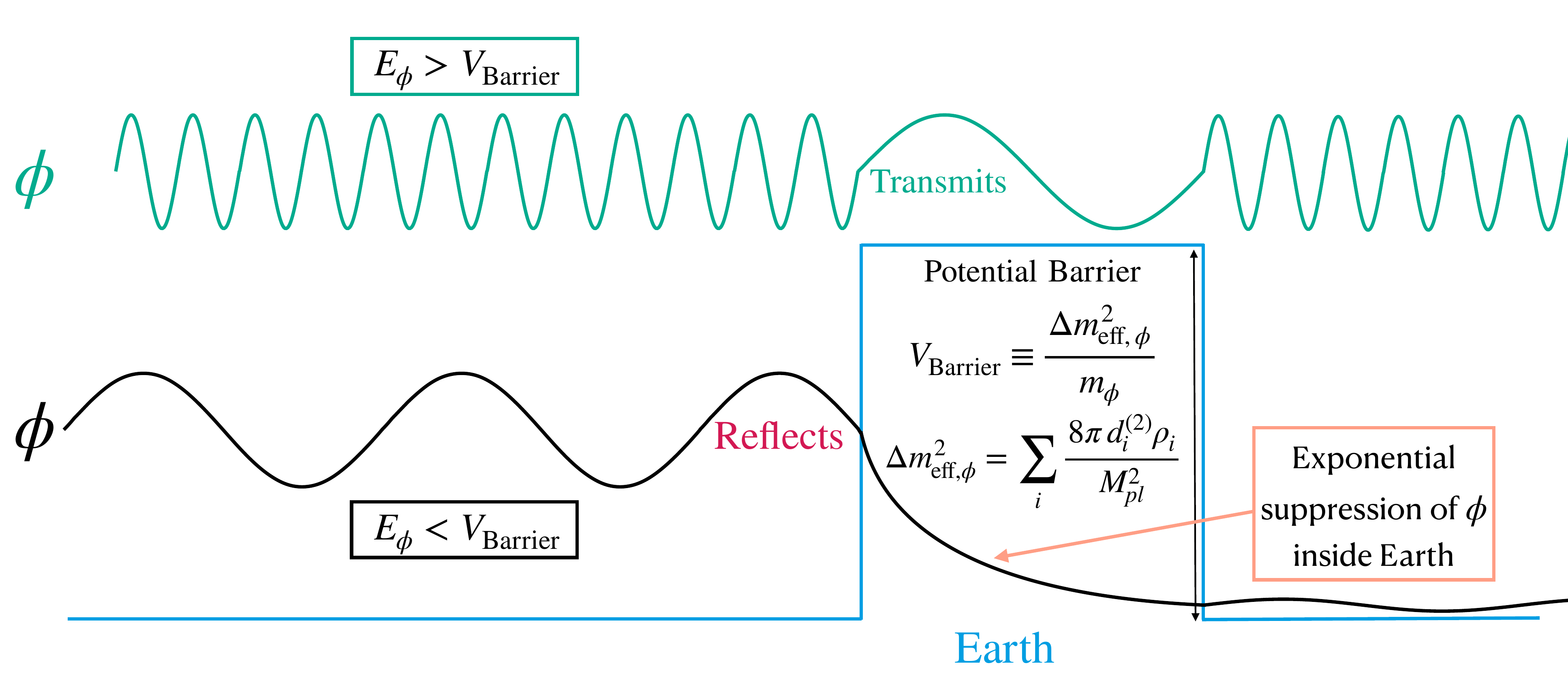}
     \caption{A pictorial representation of the Earth screening a ULDM wave, analogous to typical finite width barrier problems in quantum mechanics. Due to the quadratic $\phi-$SM coupling, the Earth represents a potential barrier, whose height is equivalent to $V_{\rm barrier} \equiv  \sum_i 8\pi\,d_i^{(2)} \rho_i/(m_{\phi}\Mpl^2)$. If the DM has kinetic energy greater than the height of the potential barrier, the DM will primarily reflect off the barrier, with an exponentially suppressed profile inside the Earth.}
    \label{fig:barrier}
\end{figure*}

A distinct consequence of quadratic ULDM couplings to SM is the screening of $\phi$ fields, arising from an increase in their effective masses in the presence of large densities of SM fields in various environments~\cite{Hees:2018fpg}. These effects can lead to significantly decreased sensitivities to such interactions in cases where critical screening occurs. In particular, for ULDM couplings proportional to $\phi^2$ the presence of SM fields modifies the effective mass of $\phi$ as
\begin{align} \label{eq:meff}
    m_{\rm eff,\phi}^2  
    = m_{\phi}^2 + \Delta m_{{\rm eff},\phi}^2 
    = m_{\phi}^2 + \sum_i \frac{8\pi\,d_i^{(2)} \langle\mathcal{O}_{\rm SM}^i\rangle}{\Mpl^2}~,
\end{align}
where 
\begin{equation}
    \langle\mathcal{O}_{\rm SM}^i\rangle \equiv \rho_i = \langle m_e \bar{e}e \rangle,~ \langle -\frac{1}{4}F_{\mu \nu}F^{\mu \nu} \rangle, ~ \langle -\frac{\beta(g_s)}{2g_s}G^a_{\mu \nu}G^{a\mu \nu} \rangle  
\end{equation} 
are the expectation values of the SM operators in the presence of electron, photon and gluon SM fields, which represent the energy density of those fields in a medium. In environments with sufficiently dense SM matter, such as the Earth, the effective mass of traversing $\phi$ can become greater than its total energy. This leads to an exponentially decaying field profile inside, such as in case of the Earth. 
As we illustrate in Fig.~\ref{fig:barrier}, the considered system can be also viewed akin to a quantum mechanical potential barrier of finite width whose potential energy is larger than the particle's kinetic energy, where the transmission probability of the field in those regions becomes significantly suppressed. 
In this case, the potential barrier height is governed by the change in effective mass between different regions, Eq.~\eqref{eq:meff}. 

The critical coupling above which 
the field $\phi$ becomes exponentially screened 
depends on the density of the environment of interest, and for terrestrial experiments particularly relevant is that of Earth. In the case where $d_{i}^{(2)}>0$, the characteristic critical coupling is given by 
\begin{align} \label{eq:di2_crit}
    d_i^{(2),\,{\rm crit}} =
        \Mpl^2 \frac{1/R^2+E_{\phi}^2 - m_{\phi}^2}{8 \pi \rho_i}\, ,
\end{align}
where $E_{\phi} = \sqrt{p_{\phi}^2 + m_{\phi}^2}$ is the total energy of $\phi$ and $R$ is the radius of the Earth, which plays the role of the potential-barrier width. 
For terrestrial experiments, the SM energy density $\rho_i \simeq 10^{-4}\rho_\oplus,10^{-3}\rho_\oplus,\rho_\oplus$ for the electron, photon, and gluon coupling respectively\footnote{These fractions can be derived from a semi-empirical formula for the distribution of mass and energy in atoms composing the Earth; see e.g.~\cite{krane1991introductory}.}, where $\rho_\oplus\simeq 1-10\,{\rm g/cm}^3$ is of order the mass density of Earth.
This sets an upper limit on the coupling that can be probed, for example, for the gluon coupling it is approximately $d_g^{(2),{\rm crit}}\simeq 10^{8}$ for $m_\phi \lesssim 10^{-14}\,{\rm eV}$. The constraint weakens at larger $m_\phi$, due to the finite-size effect becoming subdominant, i.e. when $E_{\phi}^2 - m_{\phi}^2 \equiv p_{\phi}^2 > 1/R^2$ in Eq.~\eqref{eq:di2_crit}. 
Analogously, there also exists  critical coupling associated with the dense medium of the experimental apparatus itself. However, it corresponds to a significantly larger values of the dilatonic coupling. Thus, high-precision space-based experiments sensitive to smaller couplings that are unaffected by dense terrestrial environments, are advantageous for probing broader parameter space of ULDM couplings. 

We note that dilatonic ULDM couplings with an opposite, negative, sign can in principle also appear. This is a more complicated situation with rich phenomenology, where higher-order effects are expected to be prominent and is beyond the scope of the present study. Here, we only discuss positive sign interactions and leave a thorough general investigation of other scenarios for future work~\cite{Arakawa_2024}.

\subsection{Constraints on self-interaction coupling $\lambda$}
\label{ssec:lambdaconstraints}

\subsubsection{Observational constraints}

There are a number of observational constraints on ULDM, which we summarize below and represent quantitatively in Fig.~\ref{fig:selfinteractions}. First, we consider those that arise purely from gravitational interactions and depend only on the mass $m_{\phi}$.

\begin{itemize}
    \item \underline{Lyman--$\alpha$}:
    ULDM fields can have significant de Broglie wavelengths that are relevant on astrophysical scales. Their presence can suppress cosmic structure growth below certain scales. Observations of the matter power spectrum using Lyman-$\alpha$ forest neutral hydrogen absorption seen in high-redshift quasar spectra imply that~\cite{Irsic:2017yje,Rogers:2020ltq}
    \begin{equation}
        m_\phi \gtrsim 10^{-21}\,{\rm eV}\,.
    \end{equation}
    \item \underline{Ultrafaint dwarf (UFD) galaxies}:
    wave-like fluctuations in the ULDM density field can induce anomalous heating of stellar populations~\cite{Hui:2016ltb,Church:2018sro,Bar-Or:2018pxz,Bar-Or:2020tys}. Observations of velocity dispersions in cold UFD galaxies suggest that~\cite{Dalal:2022rmp}
    \begin{equation}
        m_\phi \gtrsim 10^{-19}\,{\rm eV}\,.
    \end{equation}
    Note that this constraint depends on various assumptions, including that the DM in the outer regions of considered galaxies is not tidally stripped by galaxy mergers or encounters (see e.g.~\cite{DuttaChowdhury:2023qxg}) and that boson star formation inside does not diminish heating effects. See Ref.~\cite{Zupancic:2023qgj} for further discussion.
\end{itemize}

Additionally, there are variety of observational constraints that depend on the ULDM self-interaction coupling $\lambda$.

\begin{itemize}
    \item \underline{Bullet Cluster}: 
    the distribution of matter in the colliding Bullet Cluster of galaxies constrains the size of the DM self-interaction cross section~\cite{Markevitch:2003at}, for ULDM giving
    \begin{equation}
        \lambda \lesssim 10^{-38}\left(\frac{m_\phi}{10^{-18}\,{\rm eV}}\right)^{3/2}\,.
    \end{equation}
    
    \item \underline{Structure formation}:
    sizable attractive (repulsive) ULDM self-couplings can lead to enhancement (suppression) of the matter power spectrum on astrophysical and cosmological scales~(e.g.~\cite{Li:2013nal,Fan:2016rda}). Observations on scales of $\mathcal{O}(\textrm{Mpc})$~imply~\cite{Cembranos:2018ulm}
    \begin{equation}
        \lambda \lesssim 3\times 10^{-79}\left(\frac{m_\phi}{10^{-18}\,{\rm eV}}\right)^{4}\,.
    \end{equation} 
    Such constraints can become even stronger if the growth of structure begins earlier than the epoch of matter-radiation equality~\cite{Budker:2023sex}.
    
    \item \underline{Black-hole superradiance (BHSR)}:
    rotating black holes can spin down by copiously generating ULDM particles in a gravitational-atom-like cloud configuration. Observations of rapidly spinning black holes can thus constrain existence of such ULDM fields. Astrophysical stellar-mass black holes can constrain ULDM in the mass range $2 \times 10^{-13}\,{\rm eV} \lesssim m_\phi\lesssim 3 \times 10^{-12}\,{\rm eV}$~\cite{Arvanitaki:2010sy,Arvanitaki:2015iga,Arvanitaki:2016qwi,Baryakhtar:2020gao}, whereas supermassive black holes are sensitive to $3\times 10^{-21}\,{\rm eV}\lesssim m_\phi\lesssim 2\times 10^{-17}\,{\rm eV}$~\cite{Unal:2020jiy}. 
    Sizable ULDM self-interactions can quench the growth of superradiant instability and particle generation~\cite{Baryakhtar:2020gao,Branco:2023frw}, alleviating the constraints when
    \begin{equation}
        |\lambda| \gtrsim 10^{-84}\left(\frac{m_\phi}{10^{-18}\,{\rm eV}}\right)^{5/2}~.
    \end{equation} 

    We stress that the BHSR constraints depend sensitively on challenging astrophysical measurements and accurate determination of black hole spins, with high degree of uncertainty.
\end{itemize}

More detailed discussion of these constraints can be found in Ref.~\cite{Arakawa:2023gyq}, and references therein. In our sensitivity estimates, we use the following benchmark for the self-coupling:
\begin{align} \label{eq:lambdaB}
    \lambda_{\text{b}} = 
       \displaystyle 10^{-90} \bigg(\frac{m_{\phi}}{5\times 10^{-21}\,{\rm eV}}\bigg)^4\,,
\end{align}
which is indicated by a black dashed line in  Fig.~\ref{fig:selfinteractions}.
Given the challenges and uncertainties associated with establishing robust BHSR bounds, we do not include them in $\l_b$ here. 
In Appendix~\ref{app:resultswithBHSR} (see black dotted line in Fig.~\ref{fig:selfinteractions}) we present results for an alternative analysis with a discontinuous $\l$ benchmark, which is taken to satisfy the expected BHSR limits.
    
\subsubsection{Naturalness}

\begin{figure}
    \centering
    \includegraphics[width = .3\linewidth]{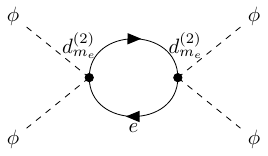}
    \hspace{.4cm}
    \includegraphics[width = .3\linewidth]{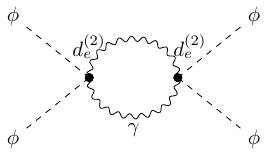}
    \hspace{.4cm}
    \includegraphics[width = .3\linewidth]{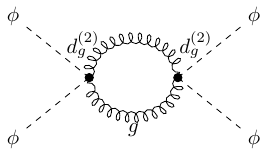}
    \caption{ One-loop diagrams that generate corrections to ULDM self-interaction coupling \(\lambda\) originating from the quadratic couplings to SM \(d_{m_e}^{(2)}\) (top left), \(d_{e}^{(2)}\) (top right) and \(d_{g}^{(2)}\) (bottom). See Eq.~\eqref{eq:Lint} of text.}
    \label{fig:quadloop}
\end{figure}

The overall $\phi^4$ self-interaction in Eq.~\eqref{eq:Lphi} is generated not only by the bare parameter $\lambda_0$, which originates in the high-energy (UV) potential of $\phi$, but also by the $\phi$-SM couplings with contributions arising at loop level; the relevant diagrams at one-loop level are shown in Fig.~\ref{fig:quadloop}. We can 
outline a limitation on \(\lambda\) motivated by naturalness, i.e. by requiring that contributions due to $\phi$-SM interactions at one-loop level are not significantly greater than the effective coupling $\lambda$, thus avoiding unnatural cancellation of these contributions with the bare Lagrangian parameter. We emphasize that the resulting upper limit on $\lambda$ will
depend on the unknown dilatonic coupling
and more generally does not constitute a strict bound. Instead, our estimates show a rough guideline for where the one-loop effects begin to become important. 

Accounting for the contribution of one-loop diagrams in Fig.~\ref{fig:quadloop} at an ultraviolet (UV) cutoff scale $\Lambda_{\rm UV}$, the quartic coupling scales as
\begin{align} \label{eq:lambdaUV}
    \lambda(\Lambda_{\text{UV}})  \simeq&~ \lambda(\mu_{0}) - \frac{(d^{(2)}_{m_e})^2 m_e^2}{4\pi^2 \Mpl^4} \Lambda_{\text{UV}}^2 + \frac{(d^{(2)}_{e})^2}{8\pi^2 \Mpl^4} \Lambda_{\text{UV}}^4 + 
    \frac{2(d^{(2)}_{g} \beta(g_s))^2}{\pi^2 g_s^2 \Mpl^4} \Lambda_{\text{UV}}^4\,,
\end{align}
where $\mu_0 \ll \Lambda_{\rm UV}$ is a reference energy scale.
The same couplings in Eq.~\eqref{eq:Lint} imply divergent contributions $\delta m_\phi^2$ to the the ULDM mass $m_\phi$ (see e.g.~\cite{Craig:2022eqo} for a recent review). Requiring a ``natural'' ULDM mass, with corrections being subdominant $\delta m_\phi \lesssim m_\phi$, implies
\begin{align}
    \Lambda^{m_e}_{\text{UV}} &\lesssim \frac{2\pi \Mpl m_{\phi}}{m_e \sqrt{d_{m_e}^{(2)}}}\,, \label{eq:Lamme}\\
    \Lambda^{e}_{\text{UV}} &\lesssim \bigg(\frac{8\pi^2 \Mpl^2 m_{\phi}^2}{d_{e}^{(2)}} \bigg)^{1/4}\,, \\
    \Lambda^{g}_{\text{UV}} &\lesssim \bigg(\frac{4\pi^2 \Mpl^2 g_s m_{\phi}^2}{\beta(g_s) d_{g}^{(2)}} \bigg)^{1/4} \label{eq:Lamg}\,.
\end{align}
Applying these limits to Eq.~\eqref{eq:lambdaUV}, the naturalness limits on renormalized self-couplings are
\begin{align}
    \lambda &\gtrsim \frac{d^{(2)}_{m_e} m_{\phi}^2}{\Mpl^2}\,, \label{eq:lamme}\\
    \lambda &\gtrsim \frac{d^{(2)}_{e} m_{\phi}^2}{\Mpl^2}\,, \\
    \lambda &\gtrsim \frac{8 d^{(2)}_{g} \beta(g_s) m_{\phi}^2}{g_s \Mpl^2} \label{eq:lamg}\,.
\end{align}
To summarize, a theory with a UV cutoff satisfying Eqs.~(\ref{eq:Lamme}-\ref{eq:Lamg}) and self-coupling satisfying Eqs.~(\ref{eq:lamme}-\ref{eq:lamg}) can be considered natural in both its quadratic (i.e. mass) and quartic (i.e. self-interaction coupling) UV contributions at one-loop order.

\begin{figure*}
    \includegraphics[width = 0.8\linewidth]{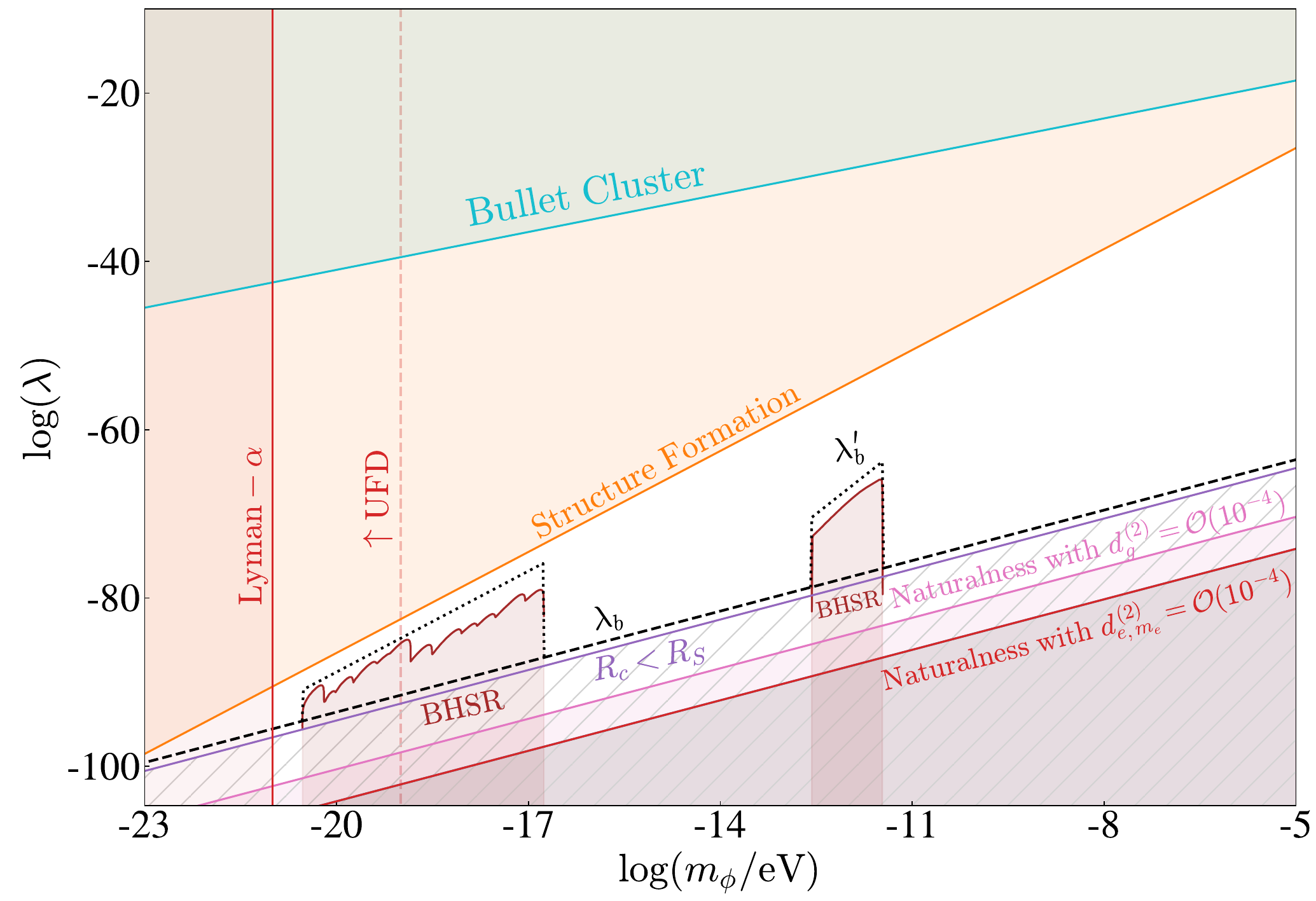}
    \caption{Constraints on ULDM self-interaction \(\lambda\) versus mass $m_{\phi}$. The bounds from the Bullet Cluster (cyan) \cite{Markevitch:2003at}, structure formation (orange) \cite{Cembranos:2018ulm}, Lyman-$\alpha$ \cite{Irsic:2017yje,Rogers:2020ltq} and ultra-faint dwarf (UFD) galaxies \cite{Dalal:2022rmp} (red vertical thick and dashed lines, respectively), black hole superradiance~\cite{Baryakhtar:2020gao,Unal:2020jiy} (BHSR, brown), as well as the lower bounds for naturalness arising from \(d^{(2)}_{m_e, e}\) (red), and \(d^{(2)}_g\) (pink) are shown. The naturalness 
    lines assume benchmark values for the couplings that are allowed by current experimental observations, as labeled. We also illustrate the region where critical boson stars would not undergo bosenova due to general relativistic effects (purple hatched region).
    The dashed black line represents the benchmark $\lambda_b$ that we choose at each mass, as given in Eq.~\eqref{eq:lambdaB} and discussed in the Sec.~\ref{ssec:lambdaconstraints}. The dotted black line shows the benchmark $\lambda_b^{\prime}$ we choose when we consider the BHSR bounds, given in Eq.~\eqref{eq:lambdaB_BHSR}. See text for details.} 
    \label{fig:selfinteractions}
\end{figure*}

The constraints from naturalness are shown in Fig.~\ref{fig:selfinteractions} for a benchmark choice of the dilatonic coupling $d_i^{(2)} = 10^{-4}$, which is allowed by all current experimental bounds. 
We note that dilatonic couplings that are larger than this are also allowed by experimental constraints at the higher end of the mass spectrum (see Sec.~\ref{sec:results} for details),
and also the dilatonic couplings could be significantly smaller.
The naturalness constraint depends on the choice of dilatonic coupling linearly, as can be seen from Eqs.~(\ref{eq:lamme}-\ref{eq:lamg}).
While we illustrate constraints from naturalness considerations, we do not consider them here as strict limits for our choice of benchmark parameter values of self-interaction coupling $\lambda$.

\section{Boson Stars and Bosenovae}
\label{sec:bosonstar}

ULDM can form gravitationally-bound objects known as \emph{boson stars}~\cite{Kaup:1968zz,Ruffini:1969qy,Colpi:1986ye}. The most important quantity characterizing a boson star is its mass $M$, which can grow in astrophysical environments, through accretion~\cite{Levkov:2018kau,Eggemeier:2019jsu,Chen:2020cef,Chan:2022bkz,Dmitriev:2023ipv} or mergers~\cite{Mundim:2010hi,Cotner:2016aaq,Schwabe:2016rze,Eby:2017xaw,Hertzberg:2020dbk,Du:2023jxh}, until it reaches a critical value~\cite{Chavanis:2011zi,Chavanis:2011zm,Eby:2014fya}
\begin{equation} \label{eq:Mcrit}
    M_c = \frac{10 \Mpl}{\sqrt{\lambda}}\,,
\end{equation}
assuming attractive self-interactions $\lambda>0$. Subsequently, the boson star collapses. Due to the minus sign on the one-loop electron correction (see Fig.~\ref{fig:quadloop} and Eq.~\ref{eq:lambdaUV}), one might consider whether or not the electron loops could contribute more during the final moments of the boson star collapse, and cause the $\lambda$ to flip sign, resulting in a respulsive self-interaction. However, since the energies associated with the bosenova are $\sim$ a few$\,\times\, m_{\phi}$, the shift in energy of the process is only a factor of a few, which is still much lower than the electron mass. We therefore do not expect this to be an important effect. In this work, we are interested in transient signals associated with such events. 

Importantly, if $\lambda$ is too small, the corresponding critical radius $R_c\simeq 0.5\Mpl\sqrt{\lambda}/m_\phi$ of a boson star would be smaller than its Schwarzschild radius $R_S=2M_c/\Mpl^2$. Hence, we will be interested only in couplings for which the critical radius satisfies $R_c>R_S$, which is equivalent to the condition
\begin{equation}
    \lambda \gtrsim \lambda_{\rm BH} \equiv 40\dfrac{m_\phi^2}{\Mpl^2}\,.
\end{equation}
This limit is illustrated by the shaded purple region in Fig.~\ref{fig:selfinteractions}. The benchmark $\lambda_b$ of Eq.~\eqref{eq:lambdaB} 
can be written as
$\l_b=10\l_{\rm BH}$.

When a boson star reaches the critical mass in Eq.~\eqref{eq:Mcrit}, it collapses gravitationally and emits $\mathcal{O}(1)$ of its mass fraction in the form of semi-relativistic ULDM particles~\cite{Eby:2016cnq,Levkov:2016rkk,Helfer:2016ljl,Eby:2017xrr}. This so-called \emph{bosenova} emission of energetic particles has already been established to constitute a novel fruitful source of signatures 
that is complementary to the conventional searches for cold DM~\cite{Eby:2021ece,Arakawa:2023gyq}.

Key properties of the bosenova emission from a boson star have been obtained with numerical simulations~\cite{Levkov:2016rkk}. Of interest for our work, the leading peak of the emission spectrum occurs around particle momentum of $\bar{k} \simeq 2.4m_\phi$ with a spread of $\delta k \simeq m_\phi$ (see also~Ref.~\cite{Eby:2015hyx,Eby:2017azn}). This first peak can be understood by energy conservation, as it is expected to be generated by 3-to-1 ($\phi\phi\phi \rightarrow \phi$) annihilation, where the initial state consists of bound state $\phi$ with energies comparable to their masses. Energy conservation then implies an energy for the outgoing $\phi$ of order $\sim 3 m_{\phi}$. The total energy emitted in this range can be estimated as~\cite{Eby:2021ece,Arakawa:2023gyq}
\begin{equation}
    \mathcal{E}_{\rm peak} \simeq 10^3\frac{m_\phi}{\lambda}\,.
\end{equation}
The emission is expected to be approximately spherically symmetric, and therefore associated ULDM bursts will travel in spherical waves from the source to the detector.

After emission, the relativistic ULDM wave will spread in-flight along the radial direction from the source. We briefly outline the consequences of these effects below, with a comprehensive discussion given in Ref.~\cite{Eby:2021ece,Arakawa:2023gyq}.~For emission sources located at astrophysical distances $r$, the wave spreading spreading in-flight rapidly comes to dominate the intrinsic burst duration $\delta t_0 \simeq 400/m_{\phi}$ \cite{Levkov:2016rkk}, which is intuitively proportional to the characteristic oscillation timescale of the bosons, $1/m_\phi$.
However, we find that given the relatively low energies of the emitted bosons, $E_\phi \simeq {\rm few}\times m_{\phi}$, the effect of wave-spreading dominates the signal duration, and the internal burst time does not significantly modify the result. Due to this, we only need to consider the signal duration due to wavespreading. This can result in a persistent signal of a significant duration at detector site
\begin{equation} \label{eq:deltat}
    \delta t \simeq \frac{\delta k}{m_\phi}\frac{r}{q^2\sqrt{q^2+1}} 
    \simeq 0.2\,{\rm year}\left(\frac{r}{\rm pc}\right)\,,
\end{equation}
where $q\equiv \bar{k}/m_\phi\simeq 2.4$ for the bosenova. 
Eq.~\eqref{eq:deltat} indicates the timescales on which the relativistic bosenova waves can interact in the detector. 
Since the signal duration is significantly protracted, one can benefit from long integration time as the burst passes through the experiment, as opposed to searching only for a signal from the leading edge of the wavefront.
The typical energy density of a bosenova at the detector site, taking into account wave spreading in-flight, is given by
\begin{equation} \label{eq:rho_*}
    \rho_* = \frac{\mathcal{E}_{\rm peak}}{4\pi r^2 \delta x}
        \simeq 3\rho_{\rm DM}
        \left(\frac{m_\phi}{10^{-15}\,{\rm eV}}\right)
        \left(\frac{10^{-80}}{\lambda}\right)
        \left(\frac{\rm pc}{r}\right)^3\,,
\end{equation}
where $\delta x\simeq \delta t$ is the spatial spread of the wave, and we have normalized to the background DM density to the local DM density near the vicinity of Earth  $\rho_{\rm DM}\simeq 0.4\,{\rm GeV/cm}^3$ (see e.g.~\cite{Read:2014qva}).

Another important consequence of the wave spreading is that the momentum modes in the wave become increasingly separated. Hence, the energy deposited in a given detector will ``chirp'' from high to low frequency over timescales $\delta t$. Although the wave is incoherent at the source, due to this spreading of momentum modes, at any given time the wavefront depositing energy in the detector can posses significant degree of coherence. The effective coherence timescale for the bosenova, after traveling a distance $r$, is given by~\cite{Eby:2021ece}
\begin{equation}
    \tau_* \simeq \frac{2\pi r}{q^3\xi}
        \simeq 4\times10^{-3}\,{\rm year}\left(\frac{r}{\rm pc}\right)\,,
\end{equation}
where $\xi \equiv m_\phi \delta t_0 \simeq 400$ is linked to the intrinsic burst duration $\delta t_0$ of the emission. This can be compared to the typical coherence time in the DM signal, $\tau_{\rm DM}\simeq 2\pi/(m_\phi v_{\rm vir}^2)\simeq 10^7/m_\phi$, where $v_{\rm vir}\simeq 10^{-3}$ is the virial velocity of the DM.

Combining the results above, for a given direct DM detection experiment we can relate its sensitivity to a bosenova signal, characterized by the coupling $d^{(2)}_{i,*}$, to its sensitivity expected for a cold DM signal, characterized by $d^{(2)}_{i,{\rm DM}}$. The ratio is given by 
\begin{equation} \label{eq:sensitivityratio}
    \frac{d^{(2)}_{i,*}}{d^{(2)}_{i,{\rm DM}}}
        \simeq \frac{\rho_{\rm DM}}{\rho_*}\frac{t_{\rm int}^{1/4}
        {\rm min}\left(\tau_{\rm DM}^{1/4},t_{\rm int}^{1/4}\right)}
        {{\rm min}\left((\delta t)^{1/4},t_{\rm int}^{1/4}\right)
        {\rm min}\left(\tau_*^{1/4},t_{\rm int}^{1/4}\right)}\,,
\end{equation}
where $t_{\rm int}$ is the livetime of the experiment, which we assume to be $t_{\rm int}=1$ year. Note that both sensitivities should be evaluated at the same frequency $E_\phi$, which will correspond to different $m_\phi$ shifted by $\bar{k}/m_\phi\simeq \mathcal{O}({\rm few})$ in this case, i.e. $E_\phi\simeq m_\phi$ for cold DM and $E_\phi = \sqrt{\bar{k}^2 + m_{\phi}^2} \simeq 2.6\,m_\phi$ for the bosenova peak described above.
When Eq.~\eqref{eq:sensitivityratio} is smaller than unity, a given experiment can probe couplings with higher sensitivity for a bosenova signal, if such occurs during the experimental livetime within a distance $r$, than to the cold DM signal.

The expected rate of bosenovae in the Galaxy is a complex topic worthy of a dedicated study beyond the scope of the present work. The predictions depend on variety of considerations including cosmological evolution~\cite{Gorghetto:2018myk,Vaquero:2018tib,Buschmann:2019icd,Gorghetto:2020qws,Buschmann:2021sdq,Saikawa:2024bta} and structure formation~\cite{Schive:2014hza,Schive:2014dra,Mocz:2017wlg,Schwabe:2020eac,Nori:2020jzx} for ULDM, as well as details of boson star formation~\cite{Levkov:2018kau,Chen:2020cef,Kirkpatrick:2020fwd,Chen:2021oot,Kirkpatrick:2021wwz}, tidal stripping~\cite{Dokuchaev:2017psd,Kavanagh:2020gcy,Edwards:2020afl},  accretion~\cite{Eggemeier:2019jsu,Chan:2022bkz,Dmitriev:2023ipv} and mergers~\cite{Mundim:2010hi,Cotner:2016aaq,Schwabe:2016rze,Eby:2017xaw,Hertzberg:2020dbk,Du:2023jxh}, among others. See also~Refs.~\cite{Eby:2021ece,Arakawa:2023gyq} for further discussion. Recent studies, under various simplifying assumptions, have estimated the DM fraction expected to consist of boson stars finding fractions of few to tens of percent \cite{Gorghetto:2024vnp, Chang:2024fol}. Additionally, Ref.~\cite{Maseizik:2024qly} found that the rate of bosenovae in a Milky Way-like galaxy can be as large as a few per day, depending on the model. Given the uncertainties and complexity of population statistics and event rate estimates, in this study, instead, we focus on characterizing signals and features of a singular bosenova event. 

We emphasize that although we focus on bosenovae events here, our analysis methodology is applicable to other potential sources of transient signals, such as instabilities of superradiant clouds around black holes or gravitational atoms around stars. Recently, it has been shown that intriguing signatures can result from accumulation of such emissions over time to comprise, in case of axions, diffuse axion background~\cite{Eby:2024mhd}.

\section{Detection}
\label{sec:detection}

\subsection{Detection methods}

A variety of distinct experiments can sensitively probe variations in fundamental constants that can result from ULDM interactions with SM, including the fine-structure constant $\alpha$.
We refer to Ref.~\cite{Antypas:2022asj} for detailed discussion of experimental probes, and below summarize the searches which are relevant to our results discussed in Sec.~\ref{sec:results}.

\begin{itemize}
    \item \underline{Atomic, molecular and nuclear clocks}: 
   Atomic, molecular, and nuclear  energy levels depend on fundamental constant values, with variations resulting in observable modification of clock frequency ratios. Comparing ratios between distinct clocks allows for unique probes of new physics interactions. Clock that are based on transitions between different electronic configurations are referred to as ``optical'' clocks. Microwave clocks are based on transitions between  atomic ground state hyperfine substates. Molecular clocks  are based on various transitions in molecules and molecular ions.  All of the clock comparisons experiments are sensitive to $d_{e}$.
    Optical clock frequency ratios are also sensitive to $d_g$ via the oscillating nuclear radius ~\cite{Banerjee:2023bjc}.
    Optical to microwave clock comparisons are sensitive to $d_{m_e}$ and $d_g$ in addition to $d_{e}$.
    
    Limits to $d_{m_e}$ include $^{133}$Cs/$^{87}$Rb atomic fountain clock frequency ratio \cite{Hees:2016gop},  frequency comparison between $^{171}$Yb optical lattice clock and $^{133}$Cs fountain microwave clock \cite{2022PhRvL.129x1301K}, and a comparison of Rb hyperfine transition  with quartz oscillator \cite{Zhang:2022ewz}. 
    
    The limits on $d_e$ come from frequency ratio measurements of $^{27}$Al$^+$, $^{171}$Yb and $^{87}$Sr optical clocks (BACON) \cite{Beloy:2020tgz}, frequency ratios of the E3 transition in a single-ion $^{171}$Yb$^+$ clock with E2 transition in the same ion and with and $^{171}$Yb$^+$ $^{87}$Sr optical lattice clock \cite{Filzinger:2023zrs}, and a dynamical decoupling demonstration in Sr$^+$ trapped ion clock transition \cite{Aharony:2019iad}. 
    
    The limits on $d_g$ come  from the $^{133}$Cs/$^{87}$Rb atomic fountain clock frequency ratio \cite{Hees:2016gop}, the limits from oscillating nuclear radius effects in $^{171}$Yb$^+$ ion clock~\cite{Banerjee:2023bjc}, and $^{171}$Yb optical lattice clock and $^{133}$Cs fountain microwave clock comparison \cite{2022PhRvL.129x1301K}. A significant improvement in $d_{m_e}$ limits is expected with the SrOH  molecular experiment \cite{SrOH2021}. 
    Very high sensitivity to variation of fundamental constants is expected of the nuclear clock \cite{2021Peik}, which rely on a nuclear transition in $^{229}$Th, leading to potential orders of magnitude improvement in $d_e$ and $d_g$ \cite{Antypas:2022asj}.
    
    \item \underline{Optical cavities}: 
    Variations of fundamental constants can induce modifications of physical dimensions of solid objects, which can be observable when sound waves propagate through objects faster than the relaxation response time. Effects of ULDM can be sensitively probed by cavity reference frequency comparisons.   H-maser comparison with a Si cavity \cite{Kennedy:2020bac} provides limits  to $d_{m_e}$ and $d_g$, comparison of Cs frequency with a cavity~\cite{Tretiak:2022ndx} is sensitive to $d_{m_e}$ and $d_e$, while frequency comparisons of strontium optical clock with silicon cavity \cite{Kennedy:2020bac}, and H-maser and sapphire oscillator with quartz oscillator~\cite{Campbell:2020fvq} are sensitive to  $d_e$. A proposal for a future cavity experiment~\cite{Geraci:2018fax} offers sensitivity to $d_{m_e}$ in a different mass range to the current optical cavity searches.

    \item \underline{Spectroscopy}: 
    Variations in fundamental constants due to ULDM-SM couplings that lead to observable signatures in atomic and molecular spectra are the basis of clock searches. However, other spectroscopy experiments can also be used for this purpose. These include molecular spectroscopy of iodine molecules \cite{Oswald:2021vtc}, frequency comparisons of $^{164}$Dy with quartz oscillator \cite{Zhang:2022ewz}, and between two isotopes of dysprosium (Dy/Dy) \cite{VanTilburg:2015oza}, which are all sensitive to $d_e$. 

    \item \underline{Mechanical resonators}: 
Both torsion balances and optical-cavity-based ULDM searches involve the measurement of the deformation of an elastic body (the mechanical response) produced by a weak force \cite{Antypas:2022asj}. The deformation can be amplified by a
factor as large as the resonance Q factor if the body
has internal resonances at the ULDM Compton frequency, yielding massive sensitivity enhancement
over a narrow bandwidth.

  Results from AURIGA \cite{Branca:2016rez} were used to put constraints on ULDM-SM coupling. Future experiments include DUAL gravitational wave detector \cite{Arvanitaki:2015iga} and compact acoustic resonators composed of superfluid Helium and crystals (Sapphire, Pillar, Quartz) \cite{Manley:2019vxy}  sensitive to both $d_{m_e}$ and $d_e$.

    \item \underline{Optical interferometers}: Modern optical interferometers can operate beyond the quantum shot-noise limit, thus offering high sensitivity to minute changes in the optical path length of their arms. Such interferometers have been used to search for and detect gravitational waves. Variations in fundamental constants due to ULDM would manifest as changes in electronic modes and lattice spacing of materials, leading to measurable change in size and refractive index of the interferometer. Using this principle, DAMNED \cite{Savalle:2020vgz}, Fermi Holometer \cite{Aiello:2021wlp}, GEO 600 \cite{Vermeulen:2021epa} and LIGO O3 \cite{Fukusumi:2023kqd} have been used to constrain ULDM couplings to electrons and photons. 
    
    \item \underline{Atom interferometers}: Atom interferometers involve splitting of matter waves and recombining them using coordinated laser pulses. 
    Particularly high sensitivity can be achieved for large-scale interferometers, prompting considerations of space-based interferometers. Here, we consider projected constraints on $d_{m_e}$ and $d_e$ from terrestrial AION and space-based AEDGE \cite{Badurina:2021rgt}, and terrestrial MAGIS \cite{2021QS&T....6d4003A} proposals. 
        
\end{itemize}

We refer to the searches above as \emph{direct DM (DDM) searches} due to the nature of their coupling directly to the DM background (see Sec.~\ref{sec:differences} for details). 
In addition to these direct searches, there are searches for ULDM emission or absorption in physical systems, which are outlined below. As we will explain in Sec.~\ref{sec:differences}, the sensitivity of such probes will scale differently with the local DM density compared to direct searches.

\begin{itemize}
    \item \underline{Equivalence-principle tests}: Ultralight scalars can induce violations in the Einstein equivalence principle (EP). Torsion balances can measure differences in forces acting on massive objects, and thus can be used to test EP violation, such as the E$\ddot{\rm{o}}$t-Wash experiment \cite{Hees:2018fpg}, which gives constraints on $d_{m_e}$ and $d_e$. The space-based MICROSCOPE experiment tests the same by measuring the difference in accelerations of two freely-falling objects of differing composition in orbit around the Earth \cite{Berge:2017ovy}, and is sensitive to $d_{m_e}$, $d_e$ and $d_g$.

    \item \underline{Fifth-force tests}: Scalar coupling to SM fields can lead to Yukawa-type interactions between matter, modifying Newton's law of universal gravitation. Tests of such ``fifth forces'' can constrain linear ULDM-SM couplings \cite{Fischbach:1996eq, Murata:2014nra}. 

    \item \underline{Astrophysical bounds}: Massive self-interacting ultralight fields contribute to the evolution of the matter power spectrum (see our previous work for more details \cite{Arakawa:2023gyq}). We have outlined several astrophysical constraints on the ULDM parameter space in Sec.~\ref{ssec:lambdaconstraints}. Constraints on photon emissivity in supernova cores from astrophysical observations can also be used to constrain ULDM-photon coupling \cite{Olive:2007aj}.

\end{itemize}

\begin{figure*}[ht]
    \includegraphics[width = \linewidth]{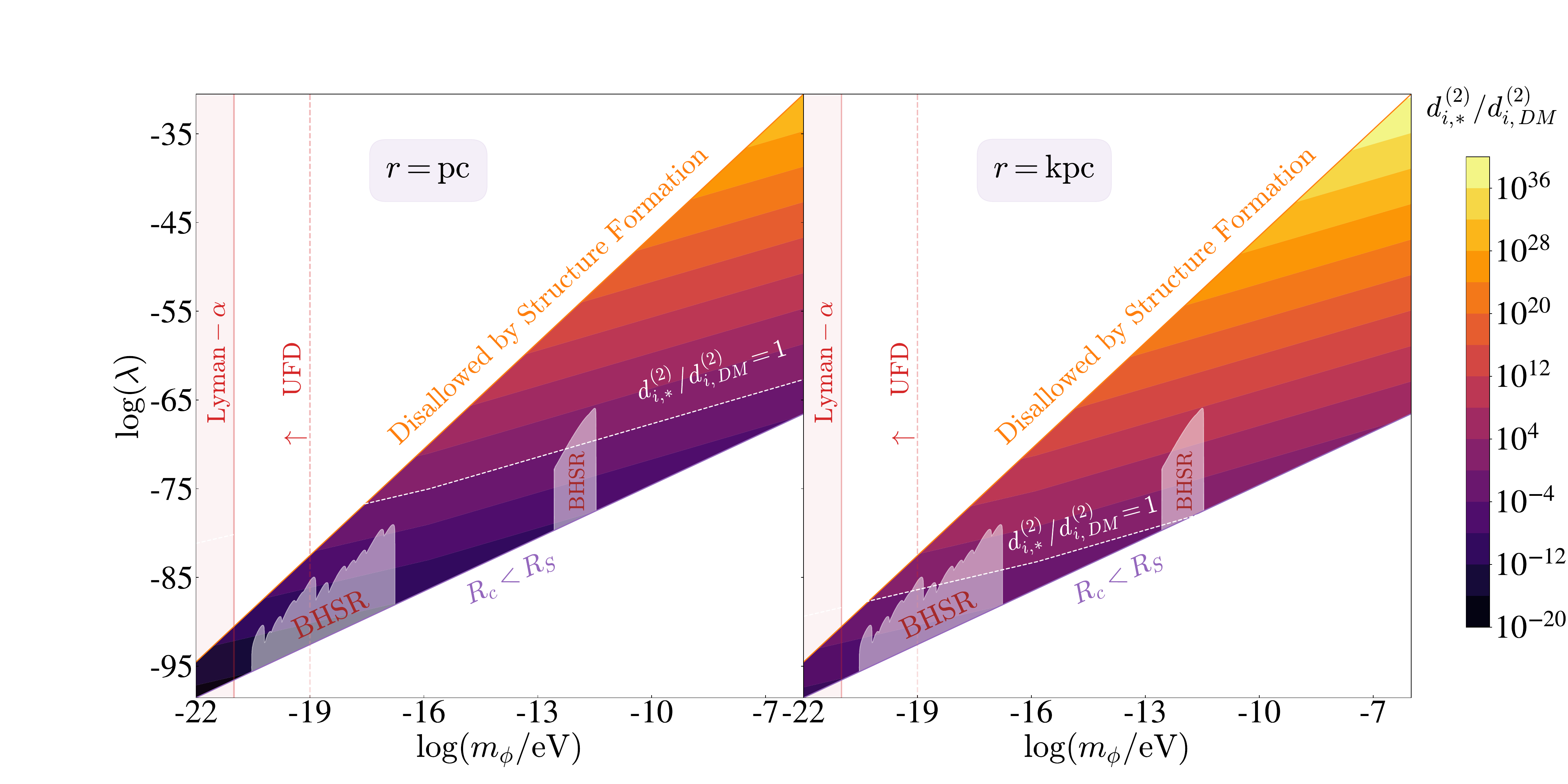}
    \caption{The sensitivity ratio of a given DM-SM coupling for a burst search $d_{i,*}$ compared to a DM search $d_i$, using Eq.~\eqref{eq:sensitivityratio}, is depicted in the allowed parameter space \(m_{\phi}\) vs \(\lambda\). The white dotted line represents equal sensitivity to bosenova and DM. Two choices for the distance to the bosenova are illustrated: \(r = \rm{pc}\) (left) and \(r = \rm{kpc}\) (right). See Sec.~\ref{ssec:lambdaconstraints} and Fig.~\ref{fig:selfinteractions} for a discussion of constraints.}
    \label{fig:gratio_contour}
\end{figure*}

\subsection{Differences with linear couplings}
\label{sec:differences}

Comparing to linear interactions, experimental sensitivities to 
fifth-forces and equivalence-principle (EP) violation are drastically modified 
for quadratic interactions.
For linear interactions, long-range forces mediated by $\phi$ arise at tree-level. However, for quadratic interactions, they are loop-level processes, and require the exchange of two $\phi$ fields at a vertex. 
The long-range effective force between the Earth and each test mass therefore depends on the background density of dark matter nearby. 

To analyze the difference quantitatively, we estimate the acceleration of test masses near the Earth as
\begin{align}
    |\vec{a}| \simeq \tilde{\alpha}(\phi) (\nabla \phi + \vec{v} \dot{\phi})\,,
\end{align}
where $\tilde{\alpha}(\phi) \propto \phi$ is a coupling function (for details, see Refs.~\cite{Banerjee:2022sqg,Hees:2018fpg, Damour:2010rm, Damour:2010rp}). The differential acceleration between two test masses is generated by the gradient of the $\phi$ profile caused by massive nearby objects. EP violation experiments seek to measure the difference in gravitational acceleration experienced by test masses with different compositions. In this case, the central body corresponds to the Earth, which generates a significant gradient. Because of this, EP tests depend on the $\phi$ background, and there exists an effective long-range force. However, fifth-force searches, search for a violation of the inverse square law between two test masses on small scales, where the gradient in the $\phi$ profile generated by the test masses is negligible. As a consequence, the fifth force searches rely purely on the quantum potential, $V_{\phi^2}(r) \sim r^{-3}$~\cite{Banerjee:2022sqg}, leading to significant suppression of their sensitivity.

The ratio of sensitivities of a given experiment to the dilatonic couplings of ultralight scalar dark matter to SM fields ($i\in(m_e, e, g)$) for the quadratic and linear cases is (see \cite{Banerjee:2022sqg})
\begin{align}
    \left(\frac{d_i^{(2)}}{d_i^{(1)}}\right)_{\rm EP} &= \frac{M_{\rm Pl} m_{\phi}}{\sqrt{\rho_{\rm DM}}}\,,\\
    \left(\frac{d_i^{(2)}}{d_i^{(1)}}\right)_{\rm DDM} &= 2^{3/2} \frac{M_{\rm Pl} m_{\phi}}{\sqrt{\rho_{\rm DM}}}\,,
\end{align}
for EP-type and direct dark matter detection experiments. 
In our results, all experimental lines scale from the linear case in either of these two ways. Astrophysical observational bounds due to Lyman-$\alpha$ forest~\cite{Irsic:2017yje,Rogers:2020ltq}, UFDs~\cite{Dalal:2022rmp} and supernova constraints on $\gamma\gamma\to\phi\phi$~\cite{Olive:2007aj} are analogous to the case of linear couplings. 

Finally, the quadratic coupling case can induce screening of the field (as discussed in Sec.~\ref{sec:EFT}), 
which is not the case for linear couplings. 
The terrestrial experiments are therefore at a disadvantage because they cannot probe arbitrarily large dilatonic couplings, and must reach sensitivities beyond the critical dilatonic coupling in Eq.~\eqref{eq:di2_crit}. All direct-detection and EP experiments that are based on the Earth are subject to this screening, which provides important motivation for future space-based searches. 

\section{Results}
\label{sec:results}

We find that bosenovae can lead to large enhancements of the ULDM density at experimental sites in a wide range of the parameter space. 
Our results complement those of Ref.~\cite{Arakawa:2023gyq} that analyzed bosenovae signatures for linear ULDM couplings to SM constituents. However, in our studied case of quadratic couplings, the signal enhancement is greater than for linear couplings. This is because the sensitivity reach of quadratic couplings follows a scaling relation of $d^{(2)}_{i,*}/d^{(2)}_{i,{\rm DM}} \sim \rho_{\rm DM}/\rho_*$, as opposed to $d^{(1)}_{i,*}/d^{(1)}_{i,{\rm DM}} \sim \sqrt{\rho_{\rm DM}/\rho_*}$ for linear couplings~\cite{Arakawa:2023gyq}. 
We demonstrate the sensitivity enhancement in the case of a bosenova search using Eq.~\eqref{eq:sensitivityratio} for the parameter space of $m_{\phi}-\lambda$ plane in Fig.~\ref{fig:gratio_contour}. 
The region below the dotted white line indicates $d_{i,*}^{(2)}/d_{i,{\rm DM}}^{(2)}<1$, implying improved sensitivity to a bosenova search provided that such a transient event occurs within distance $r$ from detector. We find that the reach for ULDM dilatonic couplings could be over 10 orders of magnitude better than that of cold DM in case of a bosenova occurring below $\lesssim$~kpc of experimental site.

\begin{figure*}
    \includegraphics[width = \linewidth]{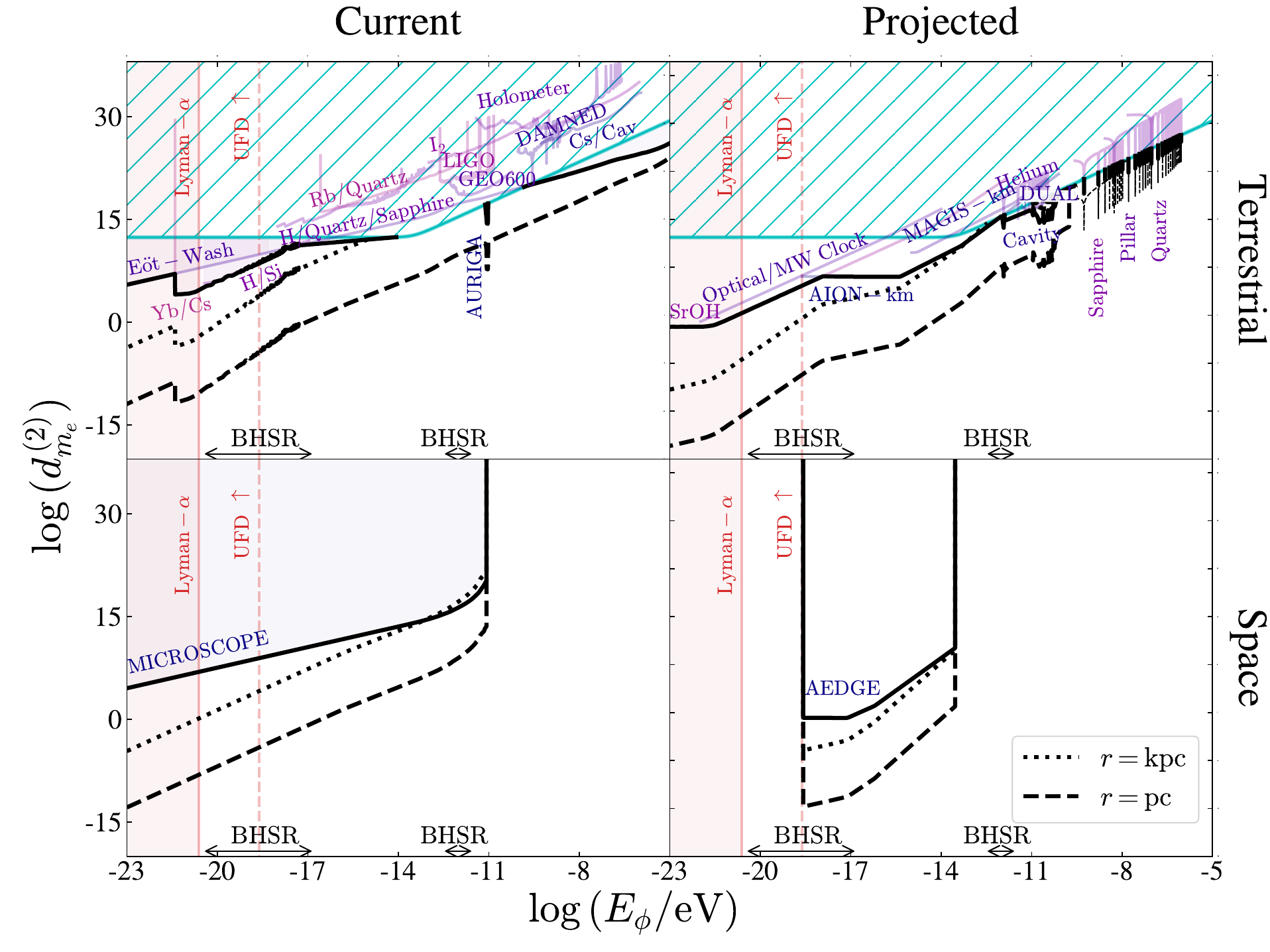}
    \caption{Current and projected  bounds on \(d_{m_e}^{(2)}\) from terrestrial and space-based experiments as a function of $E_{\phi}$. The best bounds for each mass are depicted by the solid black lines. The dashed and dotted black lines indicate the potential reach for detection of bosenovae at distances of $r = $ pc and kpc from the Earth, respectively. We choose $\lambda = \lambda_{b}$ in Eq.~\eqref{eq:lambdaB}, as explained in the text. Current experiments consist of various detector types. The following are terrestrial experiments contributing to current bounds. 
    \textbf{Clocks:} Frequency comparison between $^{171}$Yb optical lattice clock and $^{133}$Cs fountain microwave clock (Yb/Cs) \cite{2022PhRvL.129x1301K}, comparison between $^{87}$Rb hyperfine transition and quartz mechanical oscillator (Rb/Quartz) \cite{Zhang:2022ewz}; \textbf{Optical cavities:} H-maser comparison with a Si cavity (H/Si) \cite{Kennedy:2020bac}, comparison between Cs clock and a cavity (Cs/Cav) \cite{Tretiak:2022ndx}, and comparison of a H-maser and sapphire oscillator with a quartz oscillator (H/Quartz/Sapphire) \cite{Campbell:2020fvq}; 
    \textbf{Spectroscopy:} Molecular iodine spectroscopy (I$_2$) \cite{Oswald:2021vtc}; 
    \textbf{Mechanical Oscillators:} Resonant mass detector (AURIGA) \cite{Branca:2016rez}; 
    \textbf{Optical interferometers:} Unequal delay interferometer experiment (DAMNED) \cite{Savalle:2020vgz}, co-located Michelson interferometers (Holometer) \cite{Aiello:2021wlp}, GEO600 \cite{Vermeulen:2021epa}, LIGO O3 \cite{Fukusumi:2023kqd}; and 
    \textbf{Equivalence-principle tests:} E$\ddot{\rm{o}}$t-Wash \cite{Hees:2018fpg}. The only current space-based bound comes from a search for EP violation (MICROSCOPE \cite{Berge:2017ovy}).
    Future terrestrial experiments include $^{88}$Sr$^+$ optical clock and $^{133}$Cs fountain atomic clock comparison \cite{Arvanitaki:2014faa}, SrOH molecular clock \cite{SrOH2021}, optical cavities \cite{Geraci:2018fax}, resonant-mass detectors (DUAL \cite{Arvanitaki:2015iga}), other mechanical resonators (Sapphire, Pillar, Quartz and Superfluid Helium \cite{Manley:2019vxy}), and atom interferometery (AION \cite{Badurina:2021rgt}, MAGIS \cite{2021QS&T....6d4003A}). 
    Future space-based bounds come from atom interferometery (AEDGE \cite{Badurina:2021rgt}).
    Lyman-$\alpha$~\cite{Irsic:2017yje,Rogers:2020ltq} and ultra-faint dwarf galaxies (UFDs)~\cite{Dalal:2022rmp} also constrain ULDM. 
    In cyan, we show the region of the parameter space critically screened by the Earth, as discussed in Sec.~\ref{sec:Screening}.
    Finally, the regions of the parameter space corresponding to the black hole superradiance (BHSR) bounds~\cite{Baryakhtar:2020gao,Unal:2020jiy} are displayed at the bottom of the plots by the arrows. In the terrestrial plots, we also show the region of parameter space critically screened by the Earth in cyan. See Sec. ~\ref{sec:Screening} for a detailed discussion of screening of quadratically-coupled scalars.} 
    \label{fig:dme}
\end{figure*}

\begin{figure*}
    \includegraphics[width = \linewidth]{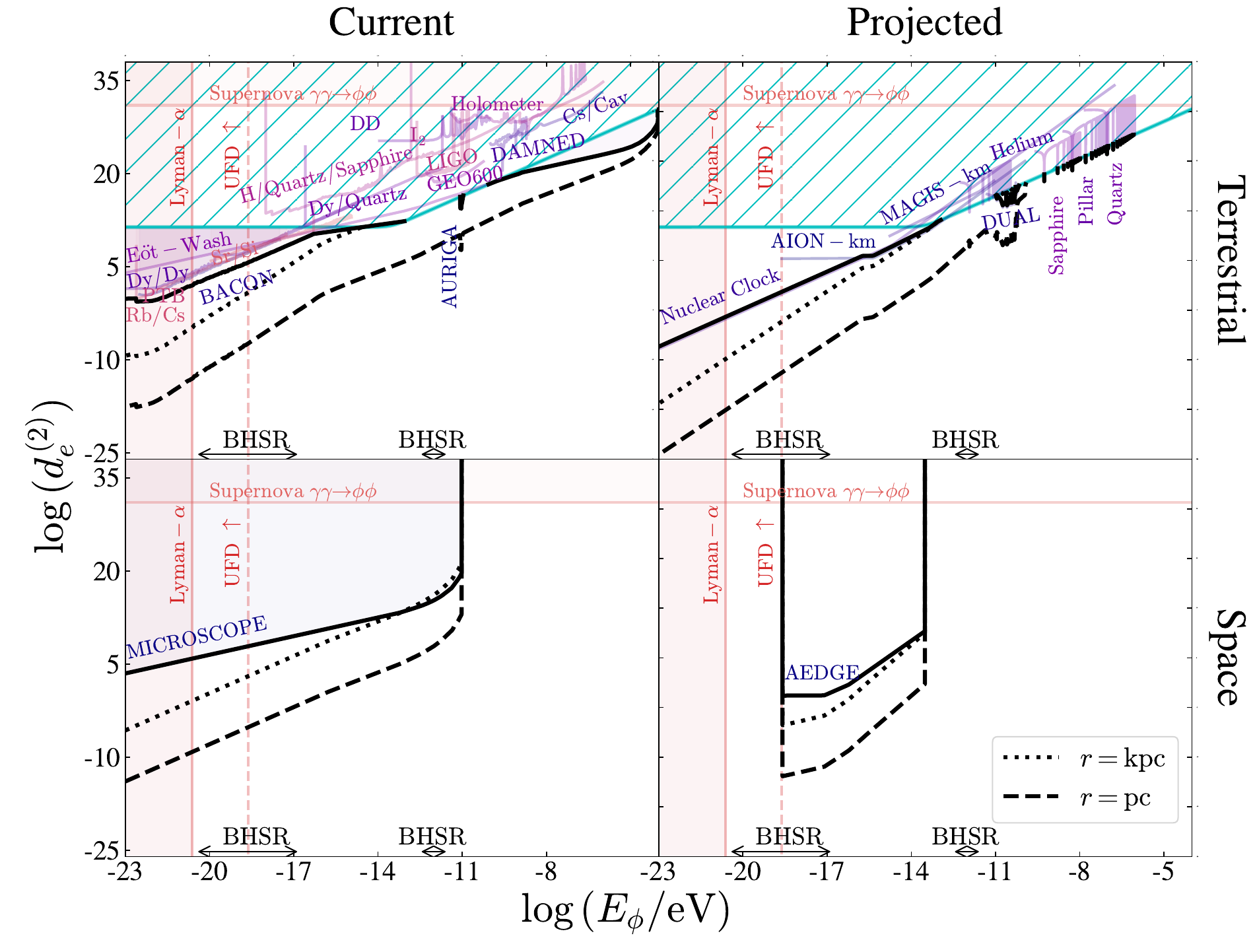}
    \caption{Current and projected bounds on \(d^{(2)}_{e}\) from terrestrial and space-based experiments as a function of $E_{\phi}$. The best bounds for each mass are depicted by the solid black lines. The dashed and dotted black lines indicate the potential reach for detection of bosenovae at distances of $r = $ pc and kpc from the Earth, respectively. We choose $\lambda = \lambda_{b}$ in Eq.~\eqref{eq:lambdaB}, as explained in the text. Current experiments consist of various detector types. The following are terrestrial experiments contributing to current bounds. 
    \textbf{Clocks:} Frequency ratios of $^{27}$Al$^+$, $^{171}$Yb and $^{87}$Sr optical clocks (BACON) \cite{Beloy:2020tgz}, frequency ratios of E2/E3 transitions in a single-ion $^{171}$Yb$^+$ clock and Yb$^+$ with $^{87}$Sr optical lattice clock (PTB) \cite{Filzinger:2023zrs}, dual $^{133}$Cs/$^{87}$Rb atomic fountain clock frequency ratio (Rb/Cs) \cite{Hees:2016gop}, and dynamical decoupling in trapped ion clock transition (DD) \cite{Aharony:2019iad}; 
    \textbf{Optical cavities:} Frequency comparisons between Strontium optical clock and Silicon cavity (Sr/Si) \cite{Kennedy:2020bac}, atomic spectroscopy in Cesium vapor with Fabry-Perot cavity locked to laser (Cs/Cav) \cite{Tretiak:2022ndx} and frequency comparison of Hydrogen maser and sapphire oscillator with quartz oscillator (H/Quartz/Sapphire) \cite{Campbell:2020fvq}; 
    \textbf{Spectroscopy:} Spectroscopic experiments of molecular iodine (I$_2$) \cite{Oswald:2021vtc}, frequency comparison of $^{164}$Dy with quartz oscillator (Dy/Quartz) \cite{Zhang:2022ewz}, precision spectroscopy measurements involving two isotopes of dysprosium (Dy/Dy) \cite{VanTilburg:2015oza}; 
    \textbf{Mechanical resonators:} Resonant mass detector (AURIGA) \cite{Branca:2016rez}; 
    \textbf{Optical interferometers:} Three-arm Mach-Zender interferometer (DAMNED) \cite{Savalle:2020vgz}, co-located Michelson interferometers (Holometer) \cite{Aiello:2021wlp}, GEO 600 \cite{Vermeulen:2021epa} and LIGO O3 \cite{Fukusumi:2023kqd};  and 
    \textbf{Equivalence-principle tests:} Test of equivalence principle violation by E$\ddot{\rm{o}}$t-Wash \cite{Hees:2018fpg}. 
    The space-based experimental bound also comes from an equivalence-principle test: MICROSCOPE \cite{Berge:2017ovy}. Future terrestrial experiments include Thorium Nuclear Clock projections \cite{Antypas:2022asj}, resonant-mass detectors DUAL \cite{Arvanitaki:2015iga}, mechanical resonators (Sapphire, Pillar, Quartz and Superfluid Helium)~\cite{Manley:2019vxy}, and atom interferometers AION \cite{Badurina:2021rgt} and MAGIS \cite{2021QS&T....6d4003A}. Future space-based bounds come from atom interferometer AEDGE \cite{Badurina:2021rgt}. In cyan, we show the region of the parameter space critically screened by the Earth, as discussed in Sec.~\ref{sec:Screening}. Lyman-$\alpha$~\cite{Irsic:2017yje,Rogers:2020ltq}, ultra-faint dwarf galaxies (UFDs)~\cite{Dalal:2022rmp}, and Supernova observations \cite{Olive:2007aj} also constrain ULDM. 
    Finally, the regions of the parameter space corresponding to the black hole superradiance (BHSR) bounds~\cite{Baryakhtar:2020gao,Unal:2020jiy} are displayed at the bottom of the plots by the arrows. In the terrestrial plots, we also show the region of parameter space critically screened by the Earth in cyan. See Sec.~\ref{sec:Screening} for a detailed discussion of screening of quadratically-coupled scalars.} 
    \label{fig:de}
\end{figure*}

\begin{figure*}
    \includegraphics[width = \linewidth]{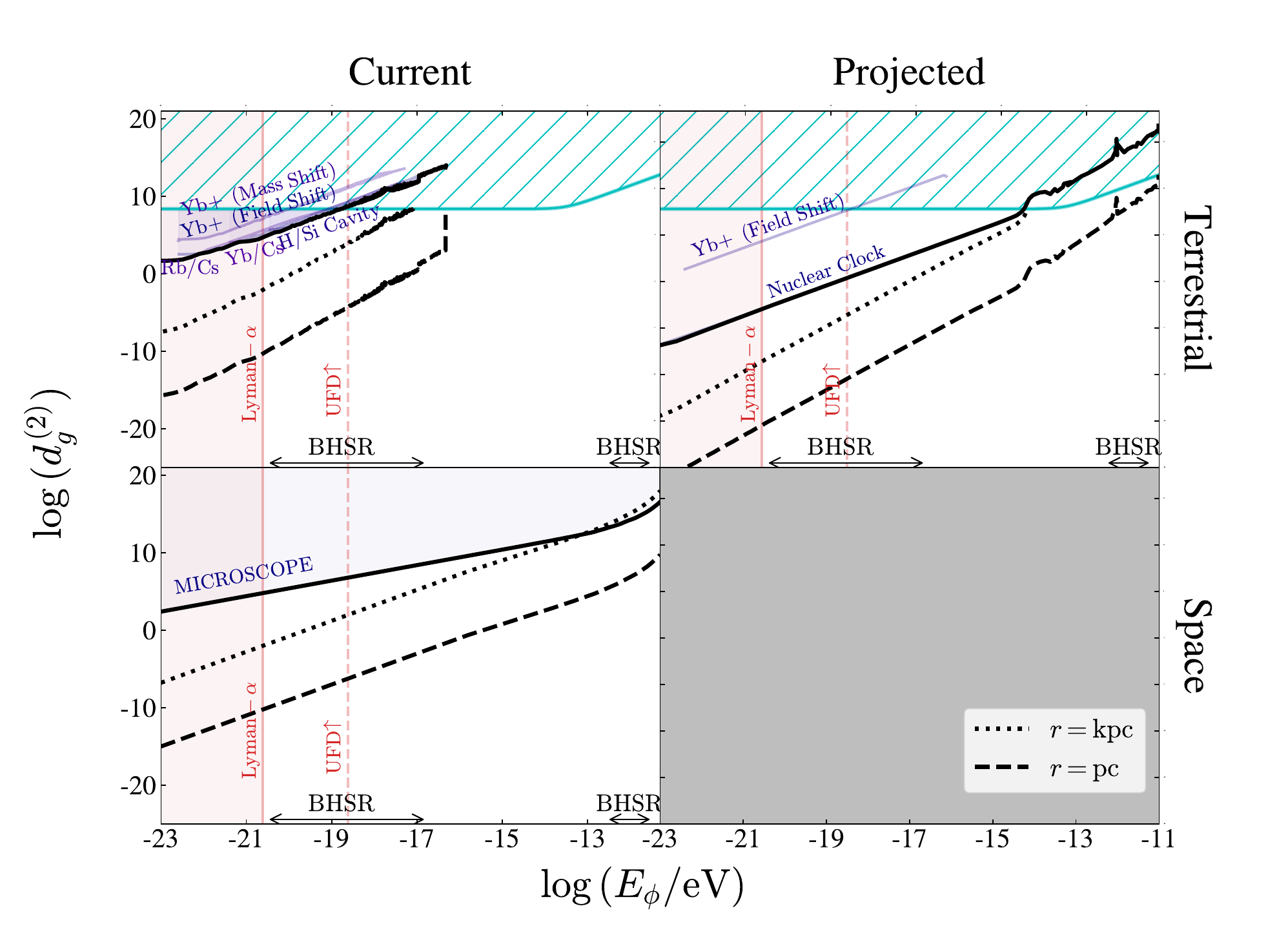}
    \caption{Current and projected  bounds on \(d^{(2)}_{g}\) from terrestrial and space-based experiments as a function of $E_{\phi}$. The best bounds for each mass are depicted by the solid black lines. The dashed and dotted black lines indicate the potential reach for detection of bosenovae at distances of $r = $ pc and kpc from the Earth, respectively. We choose $\lambda = \lambda_{b}$ in Eq.~\eqref{eq:lambdaB}, as explained in the text. The terrestrial experiments contributing to current bounds on $d_g^{(2)}$ include 
    \textbf{Clocks:} $^{171}$Yb lattice clock - $^{133}$Cs microwave clock comparison (Yb/Cs) \cite{2022PhRvL.129x1301K}, frequency comparison in dual rubidium-caesium cold atom clock (Rb/Cs) \cite{Hees:2016gop} and Yb$^+$ ion clock \cite{Banerjee:2023bjc}; and 
    \textbf{Optical cavities:} H-maser comparison with a Si cavity (H/Si) \cite{Kennedy:2020bac}; while space-based current bound comes from  
    \textbf{Equivalence-principle tests:} MICROSCOPE \cite{Berge:2017ovy}. Future experiment projections are for the $^{229}$Th Nuclear Clock \cite{Antypas:2022asj} and $^{171}$Yb$^+$ ion clock~\cite{Banerjee:2023bjc}. Note: The Nuclear Clock line in previous estimates is cut off at $10^{-21}~\rm eV$, due to the Lyman-$\alpha$ constraint. However, there is no reason that the experiment cannot be run to search for lower mass range, as in $\rm Yb^{+}$ experiment. Therefore, we extrapolate linearly to lower masses. We are not aware of any space-based planned experiments that will constrain \(d^{(2)}_{g}\).
    In cyan, we show the region of the parameter space critically screened by the Earth, as discussed in Sec.~\ref{sec:Screening}. Lyman-$\alpha$~\cite{Irsic:2017yje,Rogers:2020ltq} and ultra-faint dwarf galaxies (UFDs)~\cite{Dalal:2022rmp}  also constrain ULDM. 
    Finally, the regions of the parameter space corresponding to the black hole superradiance (BHSR) bounds~\cite{Baryakhtar:2020gao,Unal:2020jiy} are displayed at the bottom of the plots by the arrows. In the terrestrial plots, we also show the region of parameter space critically screened by the Earth in cyan. See Sec.~\ref{sec:Screening} for a detailed discussion of screening of quadratically-coupled scalars.} 
    \label{fig:dg}
\end{figure*}

In Figs.~\ref{fig:dme}, \ref{fig:de}, and \ref{fig:dg}, we display the electron ($d^{(2)}_{m_e}$), photon ($d_e^{(2)}$), and gluon ($d_g^{(2)}$) ULDM coupling reach associated with bosenova events as a function of $E_{\phi} = 2.6 \,m_{\phi}$, where the energy of the scalar field is identified with the expected from simulations leading peak in the emission spectrum from a boson star explosion~\cite{Levkov:2016rkk}. 
We showcase results for two benchmark distances between the Earth and the bosenova of $r = 1~{\rm pc}$ (dashed black) and $r = 1~{\rm kpc}$ (dotted black), considering reference coupling $\lambda$ from Eq.~(\ref{eq:lambdaB}). Each figure displays four panels, separated by the current (left column) and the projected (right column) experiments, and terrestrial (top row) as well as space-based (bottom row) experiments. The terrestrial experiments are subject to screening due to the Earth, demonstrated by the light blue hashed critical screening region, while the space-based experiments are not. Since the screening by Earth creates an upper bound on dilatonic couplings able to be probed by terrestrial experiments, we cut off the black lines if they continue above the screening region. We also represent the region where BHSR bounds would affect the results, without taking them into account for our benchmark (see Sec.~ \ref{ssec:lambdaconstraints}). We discuss the results with the BHSR constraints included in App.~\ref{app:resultswithBHSR}.

We note that the astrophysical constraints coming from Lyman-$\alpha$ and ultra-faint dwarf galaxies (UFDs) described in Sec.~\ref{ssec:lambdaconstraints} rely on 
the assumption that $\phi$ comprises the entirety of the DM abundance. Hence, they become weaker or even disappear if $\phi$ represents a sub-leading contribution to the total DM density. The laboratory bounds therefore give complementary constraints on such mass regions.


\section{Conclusions}
\label{sec:conclusions}

ULDM can form macroscopic long-lived bound boson star configurations. Their violent bosenova explosions in the presence of attractive ULDM self-interactions result in copious emissions of relativistic particles. These transient events provide intriguing novel targets for ULDM searches, with enhanced densities compared to cold DM at experimental sites that can persist for months or even years.  We find that such sources can increase the reach of current and future experiments aiming to detect ULDM by many orders of magnitude across the range of masses $10^{-23}\,{\rm eV} \lesssim m_{\phi} \lesssim 10^{-5}$ eV. This work initiates further studies that will explore detectable signatures from distinct transient sources for ULDM. 

We observe that the phenomenon of screening can significantly diminish ULDM detection prospects for sizable quadratic couplings to SM. This motivates further developments of space-based experiments that will avoid screening associated with the Earth affecting terrestrial experiments. Further, our analysis highlights that there is a ULDM parameter gap where there are no existing or planned space-based experiments that are sensitive to the larger mass range of $m_{\phi}\gtrsim 10^{-10}~{\rm eV}$.

\section*{Acknowledgements}

We thank Abhishek Banerjee, Gilad Perez, and Yevgeny Stadnik for useful discussions. 
This work was supported in part by the NSF QLCI Award OMA - 2016244, NSF Grants PHY-2012068, PHY-2309254, and the European Research Council (ERC) under the European Union’s Horizon 2020 research and innovation program (Grant Number 856415). JA thanks the International Center for Quantum-field Measurement Systems for Studies of the Universe
and Particles (QUP) for the warm hospitality and exciting research environment. The work of JE and VT was supported by the World Premier International Research Center Initiative (WPI), MEXT, Japan, and by the JSPS KAKENHI Grant Numbers 21H05451 (JE), 21K20366 (JE) and 23K13109 (VT). The work of JE was also supported by the Swedish Research Council (VR) under grants 2018-03641 and 2019-02337.
This research was supported in part by the INT's U.S. Department of Energy grant No. DE-FG02- 00ER41132. This article is also based upon work from COST Action COSMIC WISPers CA21106, supported by COST (European Cooperation in Science and Technology).

\appendix

\section{Results with Superradiance Bounds}

In the main text, 
our benchmark choice $\lambda=\lambda_b$ in Eq.~\eqref{eq:lambdaB} did not account for the existence of black hole superradiance (BHSR) constraints from rapidly-rotating black holes, discussed in Sec.~\ref{ssec:lambdaconstraints}. 
In the presence of BHSR limits, we can instead use a discontinuous benchmark of the form

\begin{align} \label{eq:lambdaB_BHSR}
    \lambda_{\text{b}}' = 
    \begin{cases}
       \displaystyle 3\times 10^{-73} \bigg(\frac{m_{\phi}}{10^{-13}\,{\rm eV}}\bigg)^6\,, 
       & \displaystyle 2 \times 10^{-13} \lesssim \frac{m_{\phi}}{\rm eV} \lesssim 3 \times 10^{-12} \\
       \\
       \displaystyle 10^{-90} \bigg(\frac{m_{\phi}}{5\times 10^{-21}\,{\rm eV}}\bigg)^4\,, 
       & \displaystyle 3\times 10^{-21} \lesssim \frac{m_{\phi}}{\rm eV} \lesssim 2 \times 10^{-17} \\
       \\
       \displaystyle 10\,\lambda_{\text{BH}}\,, & {\rm elsewhere}
    \end{cases}\,.
\end{align}

This benchmark traces the BHSR bounds, where applicable, as in the dotted line in Fig.~\ref{fig:selfinteractions}. The two discontinuities arise from observations of stellar-mass black holes in the range $m_{\phi} \simeq 3\times 10^{-13} - 3\times10^{-12}~{\rm eV}$~\cite{Baryakhtar:2020gao} and supermassive black holes in the range $m_{\phi} \simeq 3\times 10^{-21} - 2\times10^{-17}~{\rm eV}$~\cite{Unal:2020jiy}. As with Eq.~\eqref{eq:lambdaB}, this benchmark is chosen to illustrate the maximum detection prospects, since smaller values of $\lambda$ lead to larger bosenova energy density (see Eq.~\eqref{eq:rho_*}). 

\begin{figure*}
    \includegraphics[width = \linewidth]{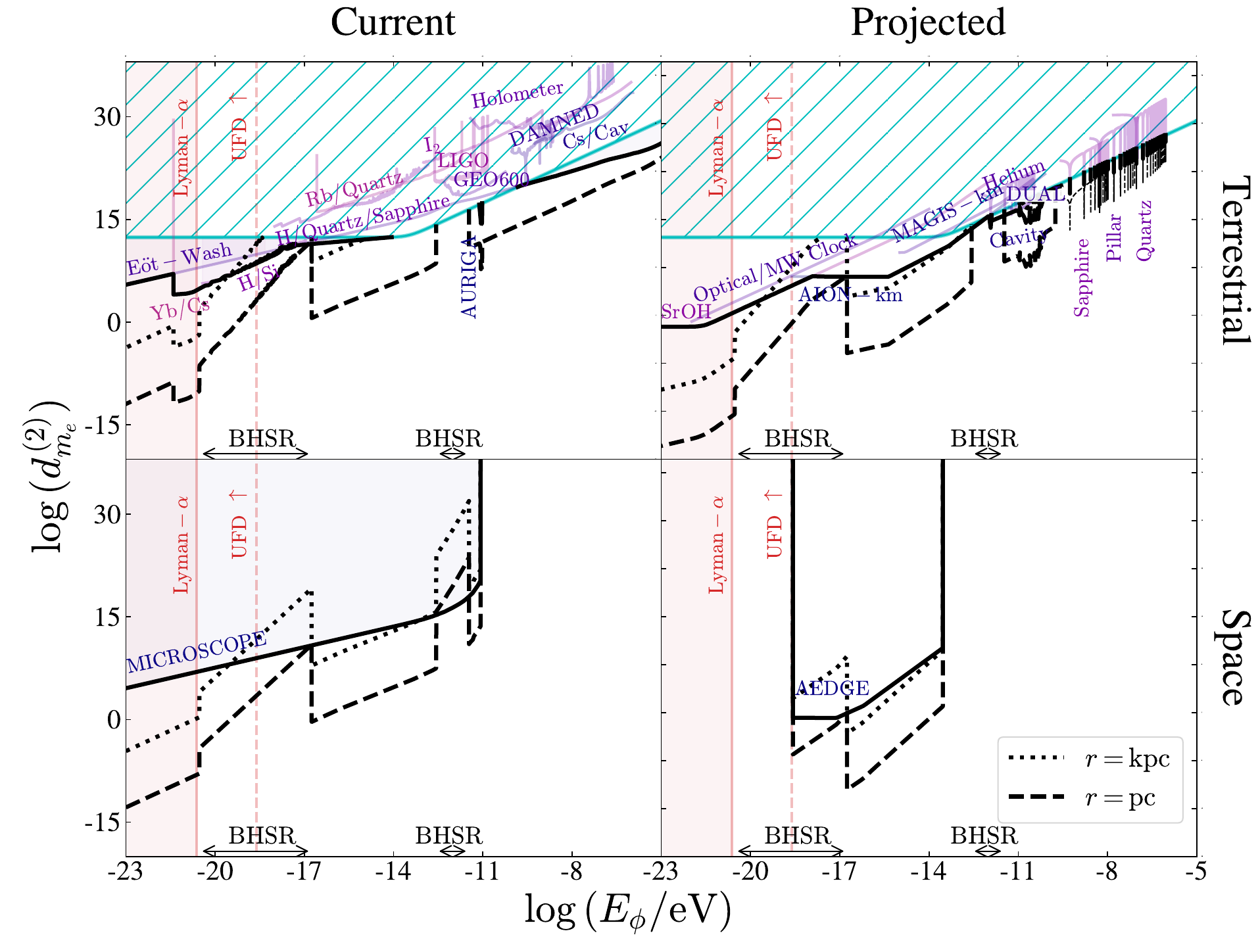}
    \caption{Current and projected  bounds on \(d_{m_e}^{(2)}\) from terrestrial and space-based experiments as a function of $E_{\phi}$ including BHSR in $\lambda$ benchmark given in Eq. \eqref{eq:lambdaB_BHSR}. The strongest limits for each mass are depicted by the solid black lines. The dashed and dotted black lines indicate the potential reach for detection of bosenovae at distances of $r = $ pc and kpc from the Earth, respectively. See Fig. \ref{fig:dme} and Sec.~\ref{sec:detection} for details and references of experimental lines.}
    \label{fig:dme_BHSR}
\end{figure*}

\begin{figure*}
    \includegraphics[width = \linewidth]{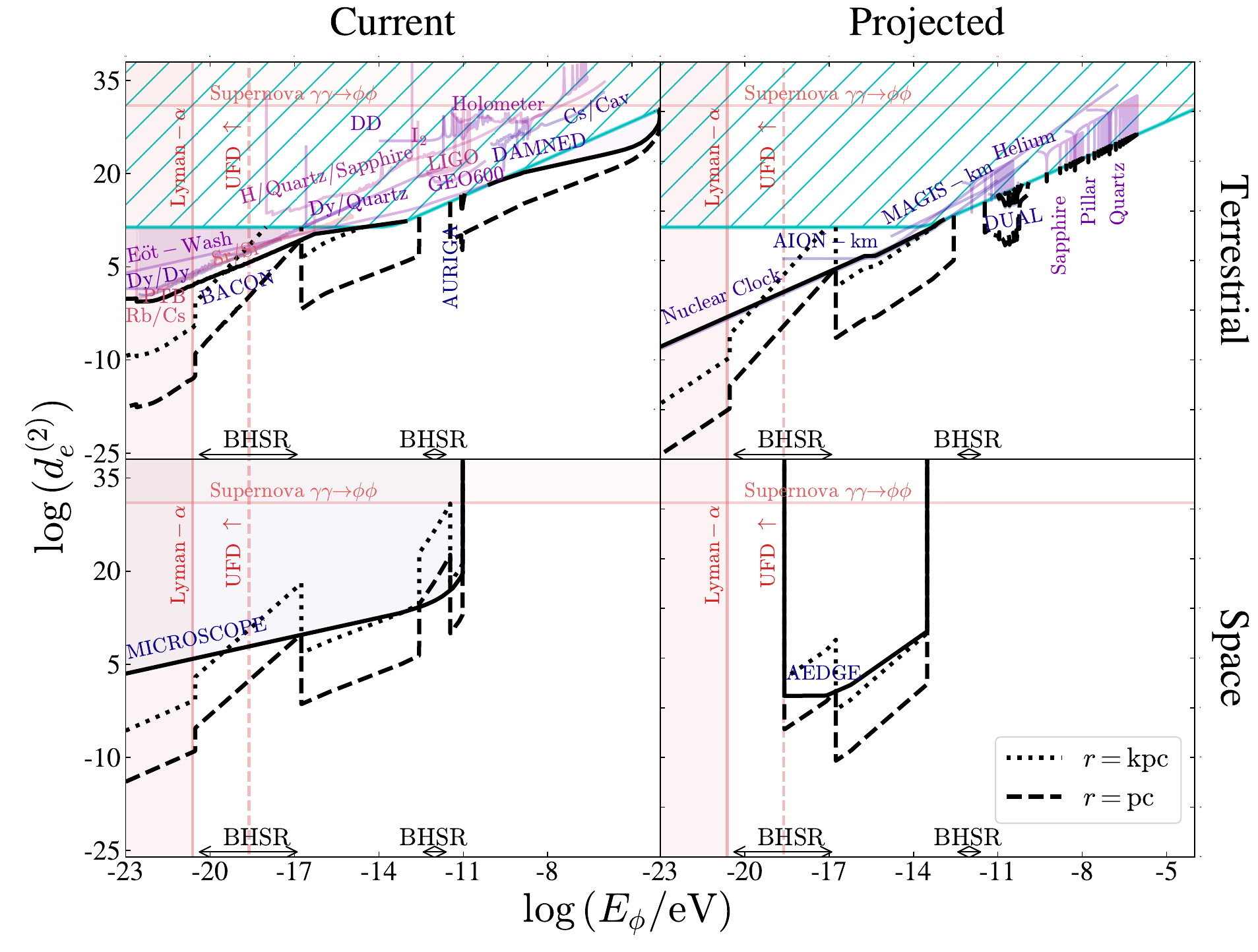}
    \caption{Current and projected  bounds on \(d_{e}^{(2)}\) from terrestrial and space-based experiments as a function of $E_{\phi}$ including BHSR in $\lambda$ benchmark given in Eq. \eqref{eq:lambdaB_BHSR}. The strongest limits for each mass are depicted by the solid black lines. The dashed and dotted black lines indicate the potential reach for detection of bosenovae at distances of $r = $ pc and kpc from the Earth, respectively. See Fig. \ref{fig:de} and Sec.~\ref{sec:detection} for details and references of experimental lines.}
    \label{fig:de_BHSR}
\end{figure*}

\begin{figure*}
    \includegraphics[width = \linewidth]{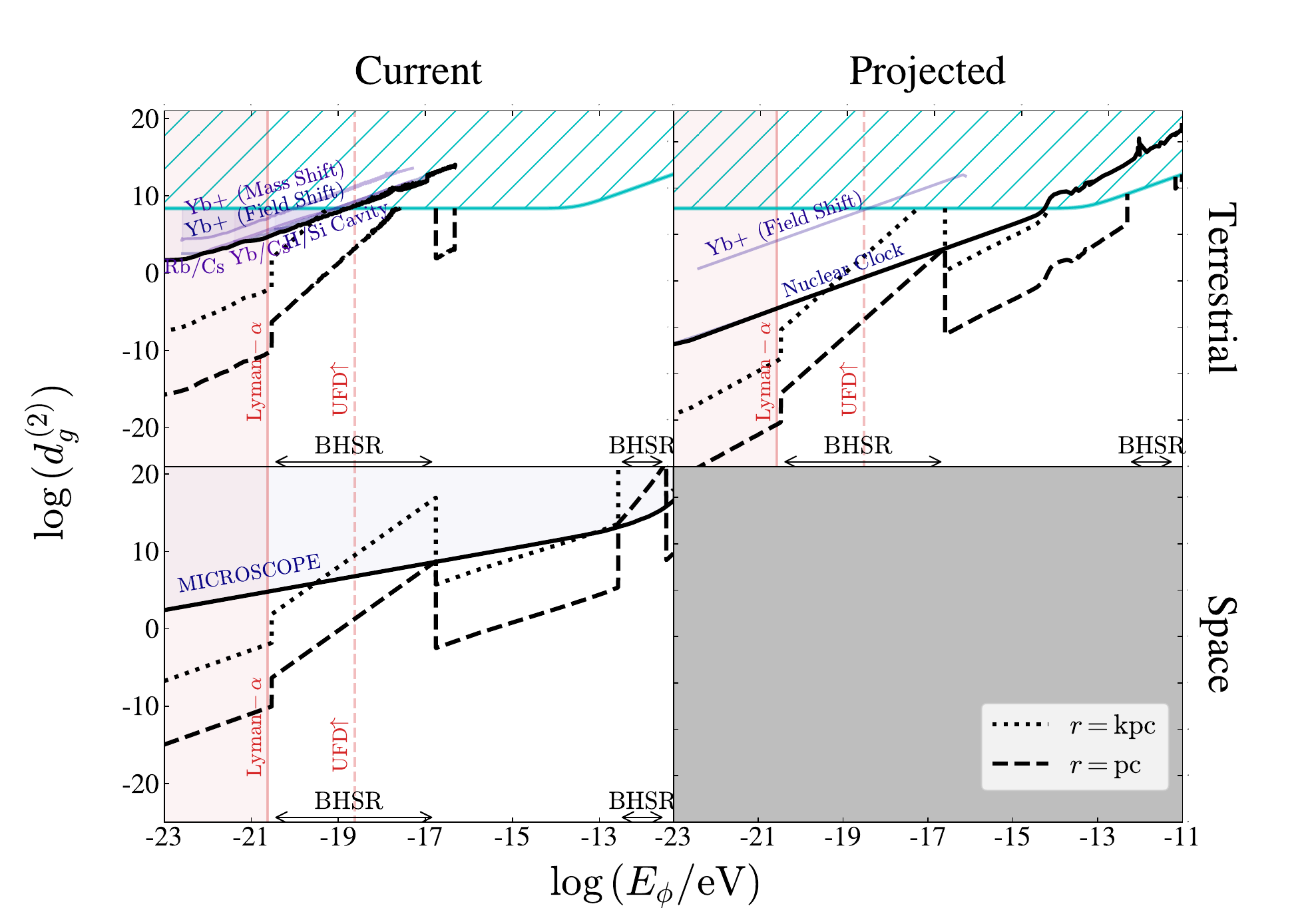}
    \caption{Current and projected  bounds on \(d_{g}^{(2)}\) from terrestrial and space-based experiments as a function of $E_{\phi}$ including BHSR in $\lambda$ benchmark given in Eq. \eqref{eq:lambdaB_BHSR}. The strongest limits for each mass are depicted by the solid black lines. The dashed and dotted black lines indicate the potential reach for detection of bosenovae at distances of $r = $ pc and kpc from the Earth, respectively. See Fig. \ref{fig:dg} and Sec.~\ref{sec:detection} for details and references of experimental lines.}
    \label{fig:dg_BHSR}
\end{figure*}

Using the modified benchmark $\lambda_b'$, we illustrate current and projected bosenovae bounds on the quadratic scalar-electron, scalar-photon and scalar-gluon ULDM couplings in Figs.~\ref{fig:dme_BHSR}, \ref{fig:de_BHSR} and \ref{fig:dg_BHSR}, respectively. 

\label{app:resultswithBHSR}

\bibliographystyle{JHEP}
\bibliography{ref}

\end{document}